\documentclass[review]{elsarticle}

\usepackage{lineno,hyperref}
\usepackage{epstopdf}
\usepackage{color}
\usepackage{mathtools}
\usepackage{amsmath}
\usepackage{amsfonts}
\usepackage{amssymb}
\usepackage{enumerate}

\usepackage{scrextend}
\usepackage{amssymb,amsmath,amsthm,enumitem}

\usepackage{listings,inconsolata}

\usepackage{comment}
\usepackage{verbatim}

\usepackage{graphicx}
\graphicspath{ {./images/} }

\usepackage{amsmath}
\usepackage{algorithm}
\usepackage[noend]{algpseudocode}

\usepackage{xcolor}

\usepackage[normalem]{ulem}

\usepackage{tikz}
\usepackage{pifont}

\usepackage{amsfonts}
\usepackage{latexsym}
\usepackage{hyperref}
\usepackage{enumerate}
\usepackage{wasysym}
\usepackage{lipsum}
\usepackage{amssymb}
\usepackage{amsthm}

\usepackage{tikz}
\usepackage{pifont}
\newcommand{\REF}[1]{(\ref{#1})}
\newtheorem{theorem}{Theorem}

\newenvironment{Reason}{\vspace{-.0em}\begin{tabbing}\hspace{2em}\= \hspace{1cm} \= \kill}
    {\end{tabbing}\vspace{-1em}}
\newcommand\Step[2] {#1 \> $\begin{array}[t]{@{}llll}#2\end{array}$ \\}
\newcommand\StepR[3] {#1 \> $\begin{array}[t]{@{}llll}#3\end{array}$
    \` {\RF \makebox[0pt][r]{\begin{tabular}[t]{r}``#2''\end{tabular}}} \\}

\newcommand\RF {\small}

\usepackage{color}

\newtheorem{example}{Example}
\newtheorem{definition}{Definition}
\newtheorem{lemma}{Lemma}

\begin{document}

\begin{frontmatter}
%\title{A Foundation for Evidence-Based Trust with Verifiable Interactions using Blockchain}
\title{A Blockchain-Based Trust Management Framework with Verifiable Interactions}

%\tnotetext[mytitlenote]{}

%% Group authors per affiliation:
%\author{Shantanu Pal, Shantanu Pal, Shantanu Pal}
%\address{Radarweg 29, Amsterdam}
%\fntext[myfootnote]{Since 1880.}

%% or include affiliations in footnotes:

%\ead[url]{www.elsevier.com}
\author[QUT]{Shantanu Pal\corref{mycorrespondingauthor}}
\cortext[mycorrespondingauthor] {Corresponding author}
\ead{shantanu.pal@qut.edu.au} 

%%\author[mysecondaryaddress]{Shantanu Pal} %% clear it
%%\author[mymainaddress]{Shantanu Pal} %% clear it
\author[UNSW,CSIRO]{Ambrose Hill}
\author[OMUCSH]{Tahiry Rabehaja}
\author[MQU]{Michael Hitchens}

\address[QUT]{School of Computer Science, Queensland University of Technology, Brisbane, QLD 4000, Australia}
\address[UNSW]{School of Computer Science and Engineering, University of New South Wales, Sydney, NSW 2052, Australia}
\address[CSIRO]{Distributed Sensing Systems Group, CSIRO's Data61, Brisbane, QLD 4069, Australia}
\address[OMUCSH]{Risk Frontiers, Sydney, NSW 2065, Australia}
\address[MQU]{Department of Computing, Macquarie University, Sydney, NSW 2109, Australia}

\begin{abstract}

There has been tremendous interest in the development of formal trust models and metrics through the use of analytics (e.g., Belief Theory and Bayesian models), logics (e.g., Epistemic and Subjective Logic) and other mathematical models. The choice of trust metric will depend on context, circumstance and user requirements and there is no single best metric for use in all circumstances. Where different users require different trust metrics to be employed the trust score calculations should still be based on all available trust evidence. Trust is normally computed using past experiences but, in practice (especially in centralised systems), the validity and accuracy of these experiences are taken for granted. In this paper, we provide a formal framework and practical blockchain-based implementation that allows independent trust providers to implement different trust metrics in a distributed manner while still allowing all trust providers to base their calculations on a common set of trust evidence. Further, our design allows experiences to be provably linked to interactions without the need for a central authority. This leads to the notion of evidence-based trust with provable interactions. Leveraging blockchain allows the trust providers to offer their services in a competitive manner, charging fees while users are provided with payments for recording experiences. Performance details of the blockchain implementation are provided.

\end{abstract}


\end{frontmatter}

%\linenumbers

%%% Introduction
\section{Introduction}
\label{itroduction}
The notion of trust is fundamental to our society. Trust provides many practical benefits including a sound and inexpensive basis for cooperation between two or more entities when uncertainty and risks exist~\cite{101007-1239199}. Commonly, a question can arise as to whether one can trust an entity and if so in what aspect and to what extent. In general, trust is used to resolve choices into decisions. Trust is context sensitive, subjective and may vary in different ways based on social and other issues~\cite{ALTAF2019}. Two entities may not draw the same conclusions about whether to trust a third entity, based on the particular context and the particular needs that confront them.

Trust in computing systems is normally evaluated based on past experience ~\cite{truong2016survey}. That is, predictions are made about an entity's future behaviour by analysing their past behaviour. When an entity is deciding whether to trust another entity, the trusting entity (or \textit{trustor}, i.e., who is trusting) may not only make use of their past interactions with the \textit{trustee} entity (i.e., who is being trusted), but may also make use of the past history of interactions that other entities have had with the trustee~\cite{JAFARIAN2020107254}. This process depends upon information about those past interactions being available. The more such evidence that is available and the more reliable such information is, the higher the quality of future predictions that can be made. This prediction normally takes the form of a trust score.

Calculations of the trust score may be based on a user's trust, device's trust or a system's trust. It may depend upon the context and the requirements of the system~\cite{GHASEMPOURI2019571}. In general, the trust score qualifies or quantifies the relationship between the \textit{trustor} and the \textit{trustee} within a specific context. This can consist of direct trust (i.e., the trustee's experience with an entity) and recommendation trust (i.e., a third party's experience with the entity). In particular, recommendation trust is useful in highly dynamic and large-scale systems to enable new interactions with some control on the risk inherent to uncertainty.  However, there is often no guarantee that the entities providing recommendation trust information have actually interacted with the trustee. If these interactions can be efficiently proven to have occurred then the basis on which trust is quantified becomes much more solid. 

The issue of trust becomes even more important given the nature and scale of present day computing systems~\cite{CHEN2021107952}. In such systems there will be no central entity which will be universally trusted to compile the evidence of interactions and calculate the degree to which entities should be trusted. Unfortunately, having such a central point is common to a number of trust system proposals. What is even more common in such proposals in the literature is a single, particular means of calculating the trust score. A multitude of such proposals has been made, varying in their approach, aims and intended context of use. On the one hand, this is understandable. As noted above, different entities will wish to make trust decisions based on different factors. On the other hand, these systems often assume all users will use the same trust calculation. We contend that such a situation needs reconsideration.

In one sense, trust can be viewed as providing information to an entity, be that entity a human user or system entity. That is, the trust system is providing a \emph{service}~\cite{1024434627}. There is nothing, at least in concept, preventing trust being regarded as a service, with different users wanting their information calculated and presented in different ways, analogous to any other service. This can also explain why users may wish to avail themselves of a trust service. As with any other service, users \emph{could} implement the functionality themselves, but it is more efficient to rely on a service provider, who, in this case, manages the transformation of evidence into a trust score. We might, indeed, imagine competing trust services, each offering to provide trust scores for entities within the system. These trust services (or trust providers) may use different algorithms for calculating trust, even charging for the service (and charging at varying rates). One issue, though, is that all the trust providers would need access to all the trust evidence. If each service attempted to collect its own set of evidence, the overall evidence base would be fragmented, leading to a lowering of the reliability of the trust scores. 

We can observe that there are three significant prerequisites for computing trust, (i) trust is relevant when there is a choice\footnote{The choice here can be binary i.e., should an entity interact with another one or not, or comparative multi-lateral i.e., which subset of entities from a larger set can be \emph{trusted}.}, (ii) trust can be built up from existing evidence, and (iii) the more reliable evidence available, the more reliable the trust score.

\begin{figure}
    \centering
    \includegraphics[scale=0.4]{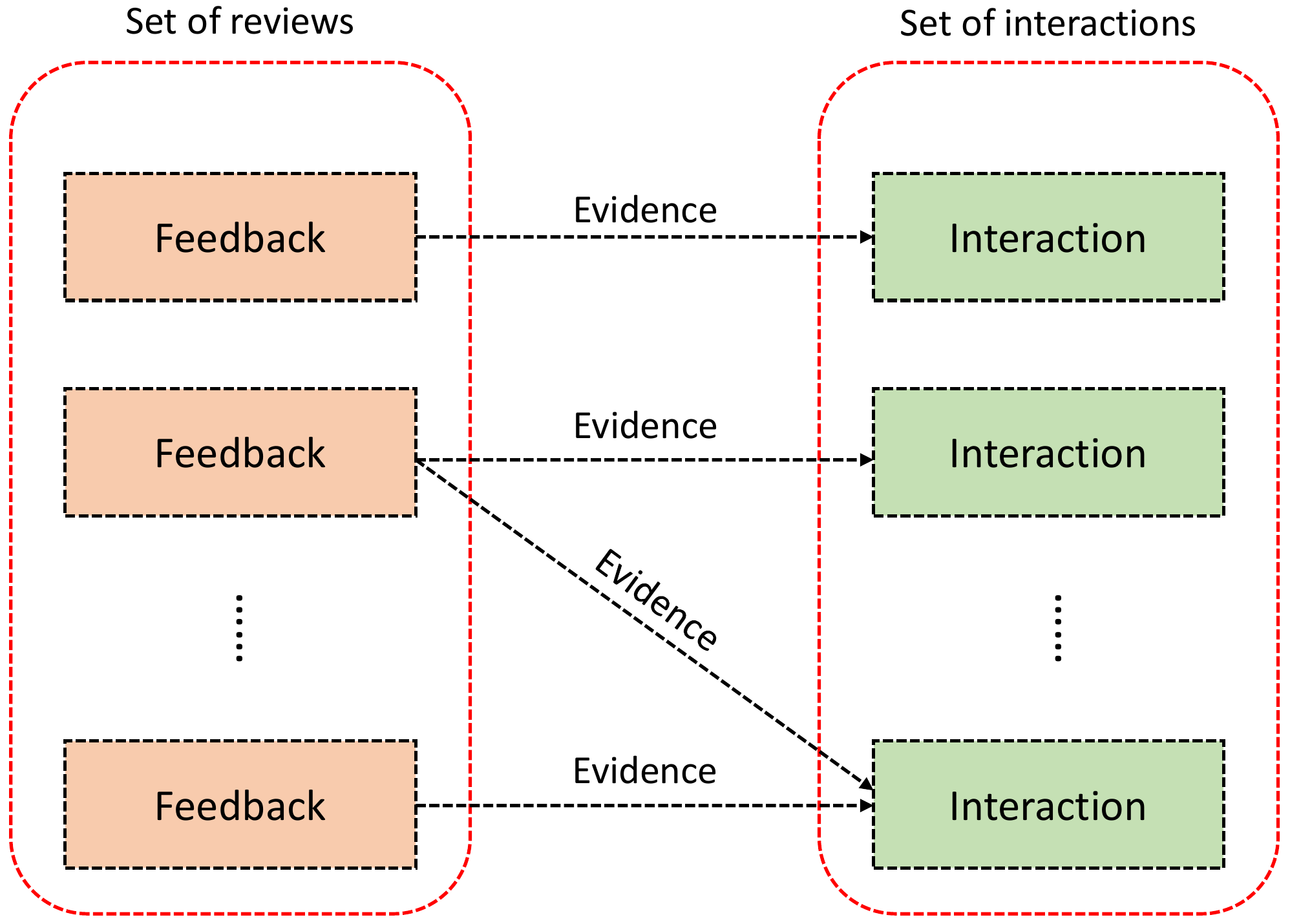}
    \caption{Relationship among feedbacks, interactions and reviews.}
    \label{fig:connection-all}
\end{figure}

In this paper, we develop a framework that enables a set of trust providers to independently provide trust scores, all using the same set of evidence. The evidence can be traced back to interactions that can be proven to have occurred. The formalisation of the trust calculation is achieved through a framework that can be tailored for specific needs. We use a global set of evidences as a universal basis for the trust calculation. An evidence is composed of an interaction and an experience in the form of a \emph{review}. If the interaction is proven to have occurred and the review is proven to pertain to that interaction, then it becomes a \emph{feedback}. Once the set of interactions and feedbacks is set, trust scores can be calculated. As an example, a basic trust scoring mechanism $\mu$ is defined on sequences of feedbacks called \emph{feedback traces}. In Figure~\ref{fig:connection-all}, we show the relationship among feedbacks, interactions and evidences. One well known example of a scoring mechanism is averaging, which takes a qualitative or quantitative average of atomic feedback ratings. A scoring mechanism can also exhibit more complex features e.g., temporal flow. The final trust score is then defined as a weighted combination of basic scores through a feedback selection $\omega$. Intuitively, each interaction may have multiple feedbacks and the feedback selection $\omega$ formally handles how much they should contribute to a particular feedback trace. We show through various examples that the combination of a scoring mechanism $\mu$ and feedback selection $\omega$ is general enough to capture many known practical cases, while providing a simple but essential formalisation. This simplicity allows us to prove important results e.g., the soundness of our framework for streams of interactions and feedbacks. In particular, we prove a convergence theorem for a broad class of scoring mechanism $\mu$ when infinite streams of feedbacks are considered.

Our formalised framework requires two fundamental properties relating to the availability of a universal set of evidences and the existence of an \emph{evidence mapping} that links feedbacks to interactions. In practice, we show that a platform using blockchain can be used to implement these two properties. A blockchain is a distributed digital ledger of transactions accessible across the network~\cite{HASSAN2019512}. In addition to being distributed, the ledger also possesses other important properties e.g., verifiable (transactions can be linked back to the involved parties, or at least their public addresses), immutable (transactions are extremely hard to amend once recorded in a block) and programmable (smart contracts can be used to automatically execute computable functions on the ledger)~\cite{8386853}. 
%
% {\color{blue}We store an interaction and feedback in this distributed ledger in the form of events. Is this still accurate Ambrose?} 
They become automatically transparent and universal and, with the help of smart contracts, we can ensure the construction of an evidence mapping. Moreover, this map can be verified, by all parties interacting through the blockchain, to be sound. That is, for each feedback, an interaction indeed occurred and the feedback is provably linked to that interaction (e.g., it was submitted by the same user that had the interaction). By storing the feedback information in the blockchain it can be accessed  by any trust provider, allowing all trust providers to base their calculation of trust scores on the same evidence set, while carrying out the calculations using different trust calculation algorithms. We can further leverage the ability of blockchains to implement financial operations to enable to provision of trust as a payable service, and even reward users for providing evidence.

We believe that the use of blockchain~\cite{BROTSIS2021108005} in this manner constitutes another step towards a robust distributed trust system that would overcome many of the challenges associated with centralisation, either in the form of a single entity calculating trust scores, or reliance on a single method for calculating such scores.
To summarize, the main contributions of this paper are as follows:

\begin{itemize}
    \item We present the foundation for evidence-based trust with verifiable interactions that can be used by any trust providers associated with the blockchain. 
    %{\color{red} This leads to the notion of trust with provable interactions.} 
    \item We develop a theoretical framework that is generic enough to be used with off-the-shelf trust scoring mechanisms and demonstrate how difference mechanisms can access the same evidence information on the blockchain.
    \item We use blockchain as a platform that implements our formalised framework. The required preconditions are ensured by the blockchain properties. 
    %Blockchains also give us additional benefits such as the direct availability of currency for interactions. 
    %\item Our proposal avoid having to scan the whole blockchain to calculate recommendation value, the total can be periodically calculated and added to the blockchain, and  when a potential trustee wants a recommendation about another entity, it looks in the blockchain.
    \item We provide a detailed proof-of-concept implementation of our proposed platform using an Ethereum network and investigate its performance. 
    \item We discuss about known attacks on trust systems and show how our proposition can improve their mitigations.
\end{itemize}

The listed contributions highlight that we are developing a trust management infrastructure rather than creating a new trust metric. We develop the notion of \emph{trust providers} which are responsible for implementing concrete and specialised trust quantification routines depending on the underlying application and the infrastructure for provision of a trust service, which allows users to select their preferred trust provider. 

\emph{Paper organisation and roadmap:} The rest of the paper is organised as follows. In Section~\ref{sec:motivation}, we present motivation of our study. In Section~\ref{preliminaries-work}, we provide preliminaries related to trust, its quantification and trust dynamics. In Section~\ref{sec-framework}, we present our framework for trust scoring with verifiable interaction. We illustrate platform design and architecture in Section~\ref{proposed-architecture}, followed by system evaluation in Section~\ref{evaluation-results}. In Section~\ref{sec:attack}, we provide a discussion on attacks on trust systems and the benefits of our proposed solution. In Section~\ref{sec-related work}, we present an extensive related work to show the novelty of our proposal over the existing proposals. Finally, we conclude the paper in Section~\ref{conclusion}.

\section{Motivation}
\label{sec:motivation}
\begin{figure}
    \centering
    \includegraphics[scale=0.35]{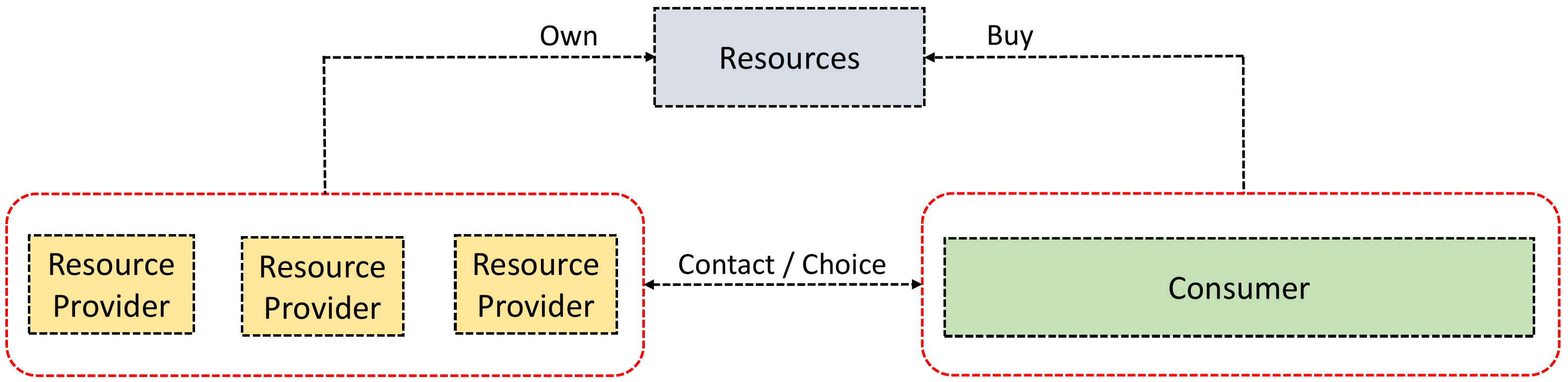}
    \caption{A simple outline of a use-case where a consumer wants to buy a resource from a resource provider.}
    \label{fig:motivation-case}
\end{figure}

In this section, we present a motivating use-scenario to show how our proposed model would work in a real-life setting. For our example use-case, we consider an Internet-based business model that is composed of many stakeholders, in our case, \emph{resource providers}, and \emph{consumers} (cf. Figure~\ref{fig:motivation-case}). A resource provider is an entity that owns a resource and wants to sell access to it. A consumer is an entity that wants to access the resource and is willing to pay for such access. When a consumer wishes to purchase an access, they need to contact a specific resource provider that manages and advertises the underlying service. That is, we are considering a digital marketplace. Further, we wish to consider situations where the resource providers and consumers are not members of a single organisation or management domain. Consumers, especially, may have varying needs and considerations that effect their preferred method of trust score calculation. That is, we are considering `open' situations where there would be no single entity trusted by all parties to collect and manage evidence and calculate trust scores.

More concretely, let us assume consumers who wish to print documents.  A range of resource providers provide printing services. The first question the consumer confronts is which resource provider to use. The services provided by different resource providers may differ. For instance, one may offer cheaper service at a lower quality (e.g., pickup from the printers themselves) while another may charge higher fees for extra convenience. In such a situation, the consumer may proceed with a choice that they themselves can resolve by investigating the service offerings and/or considering the prices offered. However, there may be service providers who offer the same or similar services that meet the consumers requirements (e.g., paper stock, print quality, etc.). In such cases the consumer is faced with a more difficult choice. The consumer can turn to a \emph{trust provider} to determine which service provider should be chosen (i.e., which is most trusted).

While a consumer will be certain of the validity of their own previous interactions with the service providers, they cannot rely on their own knowledge to provide certainty in the interactions that other consumers have had. Further, if they are to employ a trust provider, there is the question of what metric should be employed to calculate the trust score.

Previous consumers that have used the services can leave their \emph{reviews} to be viewed by others. A review becomes a \emph{feedback} once it is verified by the system that an interaction has indeed occurred. This is a significant property of our system because such verification is usually not handled by traditional trust systems --- they may simply accept the reviewer's word that an interaction has occurred. The feedbacks from the previous consumers are then ingested by a \emph{Trust Provider} and used to produce a trust score for the underlying service. The consumer can use this score to resolve their choice and decide which broker to use while taking into account all the previous interactions of their peers with the service provider.

Essentially there are two questions. Firstly, how can there be certainty in the evidence being used in calculation of the trust score and, secondly, how should the trust score be calculated. The two questions are linked, in that the calculation of the trust score depends, in part, on the evidence. Beyond that, the second question addresses the issue of what metric to employ. It cannot be said that there is a universal best trust metric, or even that in any given situation one metric is definitively superior. This is shown by the vast range of metrics proposed in the literature. Therefore different trust providers could employ different metrics, and different consumers could select between different trust providers on that basis. For example, trust metrics may differ in the weight they place on recent interactions versus interactions in the more distant, or in the means they use to combat attacks e.g., white/black washing, feedback collusion and sybil attacks. However, all trust providers need access to all the evidence to produce the most reliable trust scores.

This leads to our proposed solution, where \emph{evidence} is stored on the blockchain in the form of verified \emph{feedback}.  Such feedback can be accessed by any trust provider, which offer the provision of trust scores as a service. As an added benefit, the integration of blockchain into our system enables the payment for resource access, including trust score provision. Trust providers can implement different trust metrics, in a method analogous to different web search services implementing different search algorithms.

The notion of trust provider is one of the central components of the trust system. A trust provider uses information stored in the blockchain to derive trust scores for the resource services. This information is built from atomic blocks called \emph{evidence}. An evidence is mainly composed of two segments, (i) a \emph{Proof of Interaction} (PoI) and (ii) a \emph{feedback}. The PoI is an automatically generated token that can be linked back to a \emph{proof of access right}, another token that is generated once the consumer is given access to the printer (cf.~\cite{8894097} of Section~V). Therefore, a PoI can be linked to the actual purchase of service \emph{and} the access to the corresponding resource. The feedback is a rating that a consumer leaves based on their experience with the resource provider. Both of these data segments are stored on the blockchain which leads to many advantages discussed as follows.

\begin{itemize}
    \item The evidences from which trust scores are computed are available to all parties. This means that the set of evidence and trust scoring mechanisms has the following properties:
    \begin{itemize}
        \item \emph{Universal}: Multiple trust providers use the exact same set of evidences, even though they may use different trust scoring mechanisms producing different trust score values.
        \item \emph{Transparent}: Even though any scoring mechanism used by the trust providers may be proprietary, the underlying data used for computing the trust score is open (e.g., PageRank by Google).
        \item \emph{Verifiable/Auditable}:  An evidence can be verified by any party including the resource providers, the consumers and the trust providers. The auditability here extends to metadata. For instance, a trust provider would know for certain over what time period, with what sort of distribution the accesses and feedbacks are recorded in the blockchain.
    \end{itemize}
    \item The evidence is immutable i.e., no one can tamper with it. This applies to all the information constituting the evidence (i.e., the PoI and the feedback) and it also extends other properties e.g., logical timestamp (block number) and other metadata. Moreover, we can securely ensure that the feedback is submitted by the owner of a PoI. This is crucial for verifiable interaction.
    \item Offline trust scoring is possible. The trust calculation can be achieved offline and final scores can be stored in the blockchain. This means that the trust as a service point-of-view does not require any separate always-on component. 
    \item Review submissions can be readily incentivised using the underlying crypto-currency. Likewise, submission abuse can be penalised using the same economics.
\end{itemize}

Consumers could, in fact, use the evidence on the blockchain to carryout the trust score calculation themselves. In the same way that most users of the Internet do not implement their own search engine, we anticipate that most users would prefer to leave trust score computation to specialist trust providers.

\newcommand{\Review}{X}
\newcommand{\Interaction}{I}
\newcommand{\Service}{Y}
\newcommand{\Real}{\mathbb{R}}
\newcommand{\PositiveReal}{\mathbb{R}^+}
\newcommand{\EvidenceMap}{\varepsilon}
\newcommand{\ServiceProj}{\pi}
\newcommand{\Fcomp}{{\circ}}
\newcommand{\To}{{\to}}
\newcommand{\MapTo}{{\mapsto}}
\newcommand{\Setminus}{{\setminus}}
\newcommand{\In}{{:}}
\graphicspath{ {./images/} }
\newtheorem{innercustomthm}{Property}
\newenvironment{property}[1]
  {\renewcommand\theinnercustomthm{#1}\innercustomthm}
  {\endinnercustomthm}

\section{Preliminaries: quantitative models of trust}
\label{preliminaries-work}
In this paper, we are mostly concerned with reasoning about trust supported by recorded experience or evidence. Thus we restrict ourselves to notions of trust that can be quantified e.g., reliability, performance or quality of service. We start by briefly reviewing established quantification method using direct observation or third-party referral. We will also review trust models that are directly related to the framework we develop in Section~\ref{sec-framework}.

\subsection{Quantifying trust}
We take a quantitative approach to reason about trust, based on evidences which are built from PoI and feedback. The feedback records quantitative or qualitative measures of different aspects of the interactions that have occurred. For instance, they can represent satisfaction\footnote{Satisfaction can be directly measured through a recorded discrete value or inferred from textual feedback through sentiment analysis.}, value-to-cost ratio and so on. The feedbacks are usually personal experiences that entities experience through interacting between each other. In this case, the trust calculation is based on direct experiences by ``aggregating" individual feedback scores to form an overall personal opinion about the quality of service or reliability of a trustee. However, direct experience is not always possible since no interaction may have occurred yet between the trustor and trustee. In this case, the trustor would leverage on third-party evidences to infer a recommendation score.~\cite{DBLP:journals-5313, BELLOUSMAN2018143}.

In a recommendation, the trustor uses a referral score from one or multiple third-part that have had interacted with the trustee. For instance, user Alice wants to access a service from a service provider Bob without any prior interaction. In this situation, Alice can take a recommendation from her friend Carl who has had some interaction with Bob. This is a transitive trust calculation process where the final trust value between Alice and Bob is a combination of Alice's trust with Carl and Carl's trust with Bob. 

Since our trust calculations are based upon a universal set of evidences, the calculation of trust through direct observation and indirect recommendation is carried out through a single formal framework. This unification is important to support our notion of \emph{Trust as a Service} through the use of \emph{Trust Providers}.

\subsection{Trust models}
There are different models for quantitative trust, e.g., Belief Theory~\cite{BEYNON200037} \cite{dempster2008upper}, Subjective Logic~\cite{Josang-2001-565981}, Bayesian models~\cite{6838647, 6513398}, etc. The most relevant techniques to our approach include Dampster's Belief Theory~\cite{dempster2008upper} and Jonker and Treur's~\cite{JonkerTreur} formal approach to the dynamics of experience based trust. 
%For the Bayesian model, the interactions between the entities are defined as a binary rating. In belief model, an entity's belief about any statement is translated into rating which is further transformed into an actual trust. Finally, in case of a formal model, ...

\subsubsection{Belief theory}
This is also known as Dempster-Shafer Theory (DST) or evidence theory introduced by Dempster around 1960~\cite{dempster-upper}. It is a general probabilistic framework for reasoning about priori (incomplete) knowledge i.e., with epistemic uncertainty. Intuitively, it uses belief functions to reason about probability, possibility and plausibility. Later in the 1970s, Dempster enhanced his theory to incorporate the notion of evidence from different sources.  These evidences are collected to form a mathematical framework for reasoning about beliefs in the system. 

A central component of DST is the notion of belief and plausibility measures which are interlinked through the following equation:
\begin{equation}
\label{set-333}
\mathrm{Pl}(p) = 1-\mathrm{Bel}(\neg p)~.
\end{equation}  
Intuitively, $\mathrm{Bel}$ measures the strength of the evidence at supporting a claim expressed as a proposition $p$. Plausibility is then derived from $\mathrm{Bel}$ through Equation~\ref{set-333}. The interval $[\mathrm{Bel}(p),\mathrm{Pl}(p)]$ quantifies the epistemic uncertainty regarding proposition $p$. Our development in this paper is simpler as we assume that an evidence is restricted to a provable interaction and experience without epistemic uncertainty. In fact, the set of evidence can be verified as accurate but we leave the inherent uncertainty in ``quantifying an experience'' for future work. Thus, the scoring mechanisms we establish in Section~\ref{sec:scoring} can be noted as simple manifestations of plausibility functions.

\subsubsection{Trust dynamics through experience}
In~\cite{JonkerTreur}, Jonker and Treur analyse the dynamics of trust through a simple yet elegant formalism based on the notion of trust evolution. The core idea was to update trust based on the past evidences, i.e., a trust evolution function maps a sequence of experiences to a trust value. This update is constrained by specific properties e.g., monotonicity and positiveness/negativeness which dictate the direction of changes the evolution function should take given some recent evidence. Many other properties are also listed in~\cite{JonkerTreur}. The proposed framework in this paper can be seen as a generalisation of~\cite{JonkerTreur} with a specific focus on quantitative evaluation. We also make use of newly developed blockchain technologies to add the novel notion of verifiable interaction to this type of formalisation.

\section{Framework for trust scoring with verifiable interaction}
\label{sec-framework}
This section introduces the foundation of our trust calculation in the presence of verifiable interactions. In traditional review systems, trust scores or reputation values are mostly computed from review details e.g., score and temporal metadata. In our setting, the existence of verifiable interaction takes a central role and the notion of scoring is developed around the notion of feedback, interaction and service. In other word, feedbacks and interactions form the evidences for/or against the quality of a service.

\subsection{The fundamental maps: Evidence map and service projection}
The formalisation of our trust system is based on the evidence map $\EvidenceMap\In\Review\To\Interaction$ and the service projection $\ServiceProj\In\Interaction\To\Service$. The map $\EvidenceMap$ is a partial function from the set of reviews $\Review$ to the set of interactions $\Interaction$. Since $\EvidenceMap$ is partial, not all reviews are mapped to interactions. In contrast, any review in the domain of $\EvidenceMap$ is called a \emph{feedback}, i.e., a feedback has a proven interaction.

The projection map $\ServiceProj$ simply projects each interaction to the associated service in $\Service$. The \emph{functional composition} $\EvidenceMap\Fcomp\ServiceProj\In\Review\To\Service$ gives the set of reviews with provable interactions associated to a service. That is, for every $y\in\Service$, the set of feedbacks for $y$ is given by $(\EvidenceMap\Fcomp\ServiceProj)^{-1}(y)\subseteq\Review$~\footnote{Since the inverse function of an arbitrary map is a set function, the rigorous notation is here $(\EvidenceMap\Fcomp\ServiceProj)^{-1}(\{y\})$. However, we drop the set brackets around $y$ to make the notation lighter. It should be understood that the inverse function of a non injective function applies to sets.}.

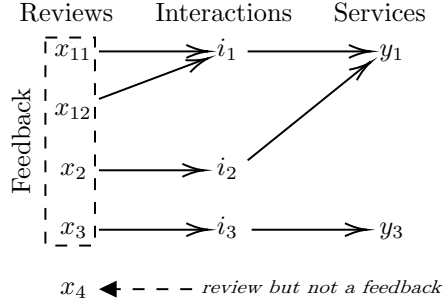
\begin{figure}[h!]
\begin{center}

\tikzset{every picture/.style={line width=0.75pt}} %set default line width to 0.75pt        

\begin{tikzpicture}[x=0.75pt,y=0.75pt,yscale=-1,xscale=1]
%uncomment if require: \path (0,300); %set diagram left start at 0, and has height of 300

%Shape: Rectangle [id:dp8556248845797955] 
\draw  [dash pattern={on 4.5pt off 4.5pt}] (96,40) -- (121,40) -- (121,145) -- (96,145) -- cycle ;

% Text Node
\draw (99,42) node [anchor=north west][inner sep=0.75pt]   [align=left] {$\displaystyle x_{11}$};
% Text Node
\draw (99,72) node [anchor=north west][inner sep=0.75pt]   [align=left] {$\displaystyle x_{12}$};
% Text Node
\draw (102,102) node [anchor=north west][inner sep=0.75pt]   [align=left] {$\displaystyle x_{2}$};
% Text Node
\draw (102,132) node [anchor=north west][inner sep=0.75pt]   [align=left] {$\displaystyle x_{3}$};
% Text Node
\draw (102,162) node [anchor=north west][inner sep=0.75pt]   [align=left] {$\displaystyle x_{4}$};
% Text Node
\draw (181,39) node [anchor=north west][inner sep=0.75pt]   [align=left] {$\displaystyle i_{1}$};
% Text Node
\draw (181,99) node [anchor=north west][inner sep=0.75pt]   [align=left] {$\displaystyle i_{2}$};
% Text Node
\draw (181,129) node [anchor=north west][inner sep=0.75pt]   [align=left] {$\displaystyle i_{3}$};
% Text Node
\draw (263,42) node [anchor=north west][inner sep=0.75pt]   [align=left] {$\displaystyle y_{1}$};
% Text Node
\draw (263,132) node [anchor=north west][inner sep=0.75pt]   [align=left] {$\displaystyle y_{3}$};
% Text Node
\draw (81.5,21) node [anchor=north west][inner sep=0.75pt]   [align=left] {Reviews};
% Text Node
\draw (150,21) node [anchor=north west][inner sep=0.75pt]   [align=left] {Interactions};
% Text Node
\draw (240,21) node [anchor=north west][inner sep=0.75pt]   [align=left] {Services};
% Text Node
\draw (83.69,92) node  [rotate=-270] [align=left] {Feedback};
% Text Node
\draw (176,161) node [anchor=north west][inner sep=0.75pt]  [font=\scriptsize,xslant=0.36] [align=left] {review but not a feedback};
% Connection
\draw    (123,47) -- (176,47) ;
\draw [shift={(178,47)}, rotate = 180] [color={rgb, 255:red, 0; green, 0; blue, 0 }  ][line width=0.75]    (10.93,-3.29) .. controls (6.95,-1.4) and (3.31,-0.3) .. (0,0) .. controls (3.31,0.3) and (6.95,1.4) .. (10.93,3.29)   ;
% Connection
\draw    (123,71) -- (176.14,51) ;
\draw [shift={(178,50.5)}, rotate = 518.71] [color={rgb, 255:red, 0; green, 0; blue, 0 }  ][line width=0.75]    (10.93,-3.29) .. controls (6.95,-1.4) and (3.31,-0.3) .. (0,0) .. controls (3.31,0.3) and (6.95,1.4) .. (10.93,3.29)   ;
% Connection
\draw    (198,102) -- (260,52) ;
\draw [shift={(260,52.5)}, rotate = 504.26] [color={rgb, 255:red, 0; green, 0; blue, 0 }  ][line width=0.75]    (10.93,-3.29) .. controls (6.95,-1.4) and (3.31,-0.3) .. (0,0) .. controls (3.31,0.3) and (6.95,1.4) .. (10.93,3.29)   ;
% Connection
\draw    (198,47) -- (258,47) ;
\draw [shift={(260,47)}, rotate = 180] [color={rgb, 255:red, 0; green, 0; blue, 0 }  ][line width=0.75]    (10.93,-3.29) .. controls (6.95,-1.4) and (3.31,-0.3) .. (0,0) .. controls (3.31,0.3) and (6.95,1.4) .. (10.93,3.29)   ;
% Connection
\draw    (123,107) -- (176,107) ;
\draw [shift={(176,107)}, rotate = 180] [color={rgb, 255:red, 0; green, 0; blue, 0 }  ][line width=0.75]    (10.93,-3.29) .. controls (6.95,-1.4) and (3.31,-0.3) .. (0,0) .. controls (3.31,0.3) and (6.95,1.4) .. (10.93,3.29)   ;
% Connection
\draw    (123,137) -- (178,137) ;
\draw [shift={(178,137)}, rotate = 180] [color={rgb, 255:red, 0; green, 0; blue, 0 }  ][line width=0.75]    (10.93,-3.29) .. controls (6.95,-1.4) and (3.31,-0.3) .. (0,0) .. controls (3.31,0.3) and (6.95,1.4) .. (10.93,3.29)   ;
% Connection
\draw    (200,137) -- (258,137) ;
\draw [shift={(258,137)}, rotate = 180] [color={rgb, 255:red, 0; green, 0; blue, 0 }  ][line width=0.75]    (10.93,-3.29) .. controls (6.95,-1.4) and (3.31,-0.3) .. (0,0) .. controls (3.31,0.3) and (6.95,1.4) .. (10.93,3.29)   ;
% Connection
\draw  [dash pattern={on 4.5pt off 4.5pt}]  (170,167) -- (126,167) ;
\draw [shift={(123,167)}, rotate = 360] [fill={rgb, 255:red, 0; green, 0; blue, 0 }  ][line width=0.08]  [draw opacity=0] (8.93,-4.29) -- (0,0) -- (8.93,4.29) -- cycle ;

\end{tikzpicture}
\caption{Evidence map and service projection.}\label{fig:evidence-map}
\end{center}
\end{figure}

An example of evidence mapping and service project is illustrated in Figure~\ref{fig:evidence-map} and explained in Example~\ref{ex:1}. In the sequel, the primitive sets $X, I$ and $Y$ are assumed to be countable, i.e., they are either finite sets or have the same cardinality as the set of natural numbers $\mathbb{N}$. 

In our trust system, the existence of the maps $\EvidenceMap$ and $\ServiceProj$ is ensured by the following two fundamental properties:

\begin{property}{1}
\label{storage}
Evidence are accessible to all parties interacting with the system. This means that the sets $X$, $I$ and $Y$ are accessible from a shared medium -- the blockchain.
\end{property}

\begin{property}{2}
\label{function}
A feedback is always supported by an interaction. 
\end{property}

In Section~\ref{sec:evaluation6.1}, we show in detail how our implementation ensures these properties using the characteristics of a blockchain.

\begin{example}
\label{ex:1}
In the system state given in Figure~\ref{fig:evidence-map}, we have a set of reviews $X = \{x_{11}, x_{12}, x_2, x_3, x_4\}$, a set of interactions $I = \{i_1, i_2, i_3\}$ and a set of services $Y=\{y_1, y_3\}$. 

\begin{itemize}
    \item The evidence map is defined by $\EvidenceMap(x_{11}) = \EvidenceMap(x_{12}) = i_1$, $\EvidenceMap(x_2) = i_2$ and $\EvidenceMap(x_3) = i_3$.
    \item The projection map is defined by $\ServiceProj(i_1) = \ServiceProj(i_2) = y_1$ and $\ServiceProj(i_3) = y_3$.
    \item The set of feedbacks is $(\EvidenceMap\Fcomp\ServiceProj)^{-1}(Y) = \{x_{11}, x_{12}, x_2, x_3\}$.
    \item $x_4$ is a review but it is not a feedback.
\end{itemize}

\end{example}

When the evidence map $\EvidenceMap$ is not injective, then we are required to ``select'' a feedback associated to the interaction to form an evidence. More generally, we define this selection using probabilistic weights as follows:

\begin{definition}
\label{def:feedback-selection}
An evidence selection is a map $\omega~\In~ \Interaction{\times}\Review\To[0,1]$ such that, for all $i\in\Interaction$, we have 
\begin{equation}
\label{eqn1}
\sum_{x\in\EvidenceMap^{-1}(i)}\omega(i,x)=1
%~\footnote{Here, we assume that the primitive set $X$ is a countable set. In general, if $X$ is a separable metric space, then $\omega:\Interaction\To\mathbb{D}X$ where $\mathbb{D}X$ is the set of probability measures over the set $X$~\cite{Part67}. The more general case would be useful to reason about continuous streams of reviews over real-time~\cite{cloudarmor}.}
\end{equation}
\end{definition}

Intuitively, for any fixed interaction $i$, an evidence selection $\omega$ chooses a feedback $x$ probabilisticaly according to the weight $\omega(i,x)$ in Equation~\REF{eqn1}. If $\EvidenceMap$ is injective, then there is only a single evidence selection map where $\omega(i,x) = 1$ iff $\EvidenceMap(x) = i$.

\begin{example}
\label{ex:evidence-selection}
Here are few common examples of evidence selections:

\begin{itemize} 

\item \emph{Deterministic selection}: for every interaction $i\in\Interaction$, there exists some feedback $x\in\EvidenceMap^{-1}(i)$ such that $\omega(i,x)=1$. This means that $\omega(i,x') = 0$ for any other review $x'\neq x$. 

\item \emph{Uniform selection}: for every $i\in\Interaction$ and $x\in\Review$, $\omega(i,x) = \frac{1}{|\EvidenceMap^{-1}(i)|}$ where $|\EvidenceMap^{-1}(i)|$ is the number of feedback in $\EvidenceMap^{-1}(i)$. This selection only applies to finite sets of feedbacks.

\item \emph{Fresh-biased selection}: for every $i\in\Interaction$ and $x_k\in\Review$ with rank $k\geq0$ out of $N$ feedbacks, $\omega(i,x_k) = \frac{(1-q)q^{N-k+1}}{1 - q^{N+1}}$ where the ratio $q\in[-1,1]$ is fixed. This selection will assign higher weights to more recent feedbacks if the rank captures the temporal order in which feedbacks are submitted. The rank can be generalised to other forms of total ordering. 

\item \emph{Geometric (decayed) selection}: for every $i\in\Interaction$ and $x_k\in\Review$ with rank $k\geq0$, $\omega(i,x_k) = (1-q)q^k$. This selection will assign higher weights to older feedbacks. This needs a $q\in[-1,1]$ and infinitely many feedbacks to converge but a finite version can also be constructed.

\end{itemize}

\end{example}

In addition to properties~\ref{storage} and \ref{function} listed above, here are few optional properties that translate directly to functional characteristics of the evidence mapping:

\begin{itemize}

\item \emph{An interaction can have at most one feedback}: this means that the function $\EvidenceMap$ is injective. This property can be implemented or simulated in many different ways depending on the underlying system. The easiest way to simulate this property is to use a \emph{deterministic} evidence selection $\omega$ which is the approach we take in this paper.

\item \emph{Any review is a feedback}: this means that $\EvidenceMap$ is a total function. This is usually hard to enforce especially on an open and distributed system like ours.

\item \emph{A feedback is mandatory after each interaction}: this means that $\EvidenceMap$ is a surjective function. Similar to the previous totality property, this is also hard to enforce in practice.

\end{itemize}

%%%
\subsection{Trust score evaluation}\label{sec:scoring}
In general, scoring functions or trust calculations are performed over sequences of feedbacks by Trust providers. In this paper, we do not assume any specific trust calculation but rather provide a general framework that supports various implementations of such calculations. We define the notion of sequence of feedbacks using the interaction set $\Interaction$ as set of indices.

\begin{definition}
\label{interaction}
A \emph{feedback trace} is a function $\alpha\In\Interaction\To\Review$ such that $\EvidenceMap\Fcomp\alpha=\mathrm{id}$ where $\mathrm{id}$ is the identity function of $I$~\footnote{Note that $\alpha$ is not necessarily total so a more precise definition is that $\EvidenceMap\Fcomp\alpha$ is a restriction of the identity function $\mathrm{id}$. However, we simply use $\mathrm{id}$ to simplify the presentation.}. A feedback trace is finite if it has a finite domain or, equivalently, range.
\end{definition}

Note this implies that a feedback trace $\alpha$ is always injective.

\begin{example}
\label{ex:feedback-traces}
Here are some feedback traces associated to the system state illustrated in Figure~\ref{fig:evidence-map}.
\begin{equation*}
\left.\begin{aligned}
&\{\}, \{i_1\MapTo x_{11}\}, \{i_1\MapTo x_{12}\}, \{i_2\MapTo x_2\}, \{i_3\MapTo x_3\}, \{i_1\MapTo x_{11}, i_2\MapTo x_2\}, \{i_1\MapTo x_{12}, i_2\MapTo x_3\}, \\
&\{i_2\MapTo x_2, i_3\MapTo x_3\}, \{i_1\MapTo x_{11}, i_2\MapTo x_2, i_3\MapTo x_3\}, \{i_1\MapTo x_{12}, i_2\MapTo x_2, i_3\MapTo x_3\},\dots
\end{aligned}\right.
\end{equation*}
\end{example}

The sequence of feedback traces in Example~\ref{ex:feedback-traces} is printed in ``increasing order". Formally, for every feedback traces $\alpha$ and $\beta$, we have 
\begin{equation}\label{feedback-ordering}
\alpha \leq \beta \textrm{ iff } \textrm{there exists $A\subseteq I$ such that } \alpha = \beta|_A
\end{equation}
This order is the well-known ordering of partial functions and it turns the set of feedback traces into a partially ordered set~\footnote{In fact, this order is directed-complete.}. A feedback trace $\alpha$ is \emph{maximal} if for every feedback trace $\beta$, $\alpha\leq\beta$ implies $\alpha=\beta$.

%\begin{lemma}\label{feedback-ordering}
%feedback traces are naturally ordered by the partial order
%\begin{equation*}
%\alpha \leq \beta \textrm{ iff } \textrm{there exists $A\subseteq I$ such that } \alpha = \beta|_A
%\end{equation*}
%\begin{proof}
%The relation $\leq$ is indeed a partial order.
%\begin{itemize}
%\item It is reflexive since $\alpha|_I = \alpha$.
%\item It is antisymmetric because if $\alpha = \beta|_A$ and $\beta = \alpha|_B$, for some $A,B\subseteq I$, then $\alpha^{-1}(X)\subseteq A\cap B$ and $\beta^{-1}(X)\subseteq A\cap B$. This means that $\alpha|_B = \alpha = \beta|_A = \beta$.
%\item It is transitive because if $\alpha = \beta|_A$ and $\beta = \gamma|_B$, then $\alpha = \gamma|_{A\cap B}$.\qedhere
%\end{itemize}
%\end{proof}
%\end{lemma}

\begin{example}
We have the following orderings between the feedback traces of Example~\ref{ex:feedback-traces}:
\begin{itemize}
\item[-] $\{\}\leq\alpha$ for every feedback trace $\alpha$,
\item[-] $\{i_1\MapTo x_{11}\} \leq \{i_1\MapTo x_{11}, i_2\MapTo x_2\} \leq \{i_1\MapTo x_{11}, i_2\MapTo x_2, i_3\MapTo x_3\}$
\item[-] $\{i_1\MapTo x_{11}\}$ and $\{i_1\MapTo x_{12}\}$ are incomparable and so is any pair of feedback traces containing them, respectively.
\item[-] $\{i_1\MapTo x_{11}, i_2\MapTo x_2, i_3\MapTo x_3\}$ and $\{i_1\MapTo x_{12}, i_2\MapTo x_2, i_3\MapTo x_3\}$ are maximal feedback traces.
\end{itemize}
\end{example}

\begin{lemma}
\label{prefix-closed}
The set of feedback traces is prefix closed. That is, if $\beta$ is a feedback trace and $\alpha\leq\beta$ then $\alpha$ is also a feedback trace.

\begin{proof}
Any restriction $\alpha|_A$ of a feedback trace $\alpha$ will also satisfy $\EvidenceMap\Fcomp(\alpha|_A) = (\EvidenceMap\Fcomp\alpha)|_A = \mathrm{id}$. The first equation follows from the fact that $\EvidenceMap\Fcomp(\alpha|_A)$ and $(\EvidenceMap\Fcomp\alpha)|_A$ have the same domain (intersection of the domain of $\alpha$ and $A$) on which they coincide.
\end{proof}
\end{lemma}

The partial order in Equation~\REF{feedback-ordering} and Lemma~\ref{prefix-closed} help ensure that every feedback trace can be obtained as the \emph{supremum} of an increasing sequence of feedback traces. This is particularly important when we are inductively building traces. Moreover, if $\EvidenceMap$ has a finite domain, then there are always maximal feedback traces with respect to the partial order $\leq$. 

Similar to sets, feedback traces also have a notion of size defined by $|\alpha| = |\alpha(I)| = |\alpha^{-1}(X)|$. It is clear to see that if $\alpha\leq\beta$ then $|\alpha|\leq|\beta|$.

\begin{definition}
\label{def:scoring-mechanism}
A \emph{scoring mechanism} is a function $\mu\In(\Interaction\To\Review)\To\PositiveReal$ such that $\mu(\alpha)\leq |\alpha|$ for every non-empty feedback trace $\alpha$. 
\end{definition}

 Intuitively, a scoring mechanism maps a feedback trace into a non-negative score. It does not need to be total but it must be defined over all possible feedback traces, which form a subset of $\Interaction\To\Review$. 

 The condition $\mu(\alpha)\leq |\alpha|$ ensures that a score can only grow as fast as the feedback trace, which is a natural assumption. More precisely, this condition ensures that $\mu$ is uniformly bounded even though there can be infinitely many reviews associated to finitely many interactions. A slightly more general assumption can be written as $\mu(\alpha)\leq C|\alpha|$ where $C$ is a fixed constant independent of $\alpha$. For instance, $C$ can be the absolute maximum rating a feedback can have. The value of the constant $C$ can vary from a scoring mechanism to another but it is not important for our theoretical development. Thus, we assume $C = 1$ for all scoring mechanism. 

A scoring mechanism takes a special value on the empty feedback trace $\{\}$. For instance, if $\mu$ is restricted to have values in $[0,1]$ then a convention $\mu(\{\}) = 0.5$ means that, in the absence of any evidence, we assign a fair score of $0.5$. Such a convention is entirely left to the implementation and can vary from one trust provider to another.

We now give the definition of trust score in the context of verifiable interactions.

\begin{definition}
\label{scoring}
Let $A$ be a finite set of interactions. The \emph{$A$-recommendation score} with respect to a scoring mechanism $\mu$ and an evidence selection $\omega$ is defined by
\begin{equation}
\label{eqn:scoring}
\sigma(A) = \sum_{\substack{\EvidenceMap\Fcomp\alpha=\mathrm{id}\\ \alpha^{-1}(X)=A}} \mu(\alpha)\prod_{\substack{\alpha(i)=x}}\omega(i,x)
\end{equation}
\end{definition}

Intuitively, the sum in Equation~\REF{eqn:scoring} is over the maximal elements of the set of feedback traces with domain included in $A$. The generic scoring $\mu$ evaluates these maximal feedback traces. Each of the maximal feedback trace is associated with a weight accumulated over the interactions through the feedback selection $\omega$. For a given feedback trace $\alpha$, this weight is $\prod_{\alpha(i)=x}\omega(i,x)$. Thus, the final score $\sigma(A)$ is the weighted average of the scores of each feedback trace with domain $A$.

Definition~\ref{scoring} is very general and facilitates the proofs for our technical foundation. A finite context $A$ is important for the sum on the right hand side of Equation~\REF{eqn:scoring} to be well-defined. The sum can still be over infinitely many feedback traces but we prove in Theorem~\ref{thm:convergence} that the right hand side of Equation~\REF{eqn:scoring} always converges if $A$ is a finite set. If $A$ is infinite countable then there is a sequence of finite sets $(A_n)_n$ such that $A_n\subseteq A_{n+1}$ and $\bigcup_nA_n = A$. In this case, we define
\begin{equation}
\sigma(A) = \lim_{n\To\infty}\sigma(A_n)
\end{equation}
when the right hand side exists. In contrast, $\sigma(\emptyset) = \mu(\{\})$, a score chosen by the underlying trust provider.

In practice, we need to compute the recommended trust score of a service. Definition~\ref{scoring} then specialises as follows.
\begin{definition}
\label{scoring-service}
Let $y\in Y$ be a service and $U$ be a finite set of interactions. The \emph{$U$-recommendation score} for service $y$ with respect to $\mu$ and $\omega$ is 
\begin{equation}
\label{eqn:scoring-service}
\sigma(y|U) = \sigma(\ServiceProj^{-1}(y)\cap U)
\end{equation} 
\end{definition}

In practice, the context $U\subseteq\Interaction$ in Definition~\ref{scoring-service} is obtained by looking at the interactions belonging to a set of users. For instance, a \emph{recommendation trust score} from a given user is obtained by setting $U$ to be the set of interactions that user has had dealing with service $y$. If the trust scoring is user agnostic, that is $U=\Interaction$, then we simply write 
\begin{equation}\label{eq:1431}
\sigma(y|I) = \sigma(y) = \sigma(\pi^{-1}(y))
\end{equation}
Moreover, if $\EvidenceMap$ is injective, then the inverse function $\EvidenceMap^{-1}$ exists and is the only maximal feedback trace. That is, any feedback $\alpha$ satisfies $\alpha\leq\EvidenceMap^{-1}$ making $\EvidenceMap{-1}$ the \emph{maximum} element of the set of possible feedback traces. In this case, the only feedback trace involved in the sum for $\sigma(A)$ in Equation~\REF{eqn:scoring} is $\EvidenceMap^{-1}|_A$. Thus Equation~\REF{eq:1431} gives
\begin{equation}
\sigma(y) = \sigma(\pi^{-1}(y)) =  \mu\left(\EvidenceMap^{-1}|_{\ServiceProj^{-1}(y)}\right)
\end{equation}
where the second equation follow from the fact that $\EvidenceMap$ is injective and we only have deterministic feedback selection.
Here, $\EvidenceMap^{-1}|_{\ServiceProj^{-1}(y)}$ is the restriction of $\EvidenceMap^{-1}$ to the set of interactions $\ServiceProj^{-1}(y)$.

The function $\sigma$, or alternatively the pair $(\mu,\omega)$, is a scoring function that is implemented by each trust provider. Each scoring function can have its own characteristics that mitigates partially or fully against certain type of attacks (see Section~\ref{sec:attack}). 

In general, a naive implementation of the recommendation score function $\sigma$ can be exponential in the number of interactions. More precisely, if there are $m$ interactions with $n$ feedbacks each, then a direct implementation of Equation~\REF{eqn:scoring} would have a time complexity of $O(n^m)$. Note however that optimised implementations can be achieved. For instance, the average scoring below can have an \emph{online complexity} of $O(1)$ if the evidence selection is deterministic and only the current average score and the number of past interactions are stored. The characterisation of the efficient scoring mechanisms is out of the scope of this paper.

\subsection{Average score}
One of the most common form of scoring mechanism is averaging. Let us assume the existence of a binary satisfaction rating function $\rho\In\Review\To\{0, 1\}$ which projects a review to its recorded rating --- either good $(1)$ or bad $(0)$. The average scoring mechanism $\mu$ is defined by 
\begin{equation}
\mu(\alpha) = \frac{\sum_{i} \rho\Fcomp\alpha(i)}{|\alpha(I)|}~~\footnote{Sums like this, i.e., without explicit restriction on $i$ in the sum, is understood to range over the domain the summed function. In this case, the sum is over the domain of $\alpha$.}
\end{equation}
Averaging is quite robust in the sense that, when there is a large amount of feedback, the score is stable and a large amount of effort is required to generate a considerable score change. However, it also has its drawbacks. For instance, it ignores the temporal information associated with the order in which feedbacks were submitted.

We can define a scoring mechanism which takes advantage of the temporal sequence of interactions as in Example~\ref{ex:evidence-selection}. In fact, $\mu$ can be extremely complex as the interactions and feedbacks have, at least, the following metadata: \emph{user identification, service identification, access identification, real or logical time} and \emph{service rating}. In practice, there should be a balance between robustness, complexity and efficiency.

\begin{example}
Let us consider the system state depicted in Figure~\ref{fig:evidence-map} and a trust provider using the average scoring mechanism and uniform evidence selection. Assume the ratings associated to each respective feedbacks are $\rho(x_{11}) = 0$ and $\rho(x_{12}) = 1 = \rho(x_2)$. We have the following evaluation and properties.
\begin{itemize}
    \item For service $y_2$, $\sigma(y_2, \{i_1\}) = 0$ but $\sigma(y_2, \{i_2\}) = \sigma(y_2) = 1$.
    \item For service $y_1$, the are two possible maximal feedback traces: $\alpha_1 = (i_1{\MapTo}x_{11}, i_2{\MapTo}x_2)$ and $\alpha_2 = (i_1{\MapTo}x_{12}, i_2{\MapTo}x_2)$. We have $\mu(\alpha_1) = \frac{\rho(x_{11}) + \rho(x_2)}{2} = 0.5$ and $\mu(\alpha_2) = \frac{\rho(x_{12}) + \rho(x_2)}{2} = 1$. Since $\omega$ is uniform, the overall score of service $y$ is 
    \[
    \sigma(y_1) = \mu(\alpha_1){\times}0.5 + \mu(\alpha_2){\times}0.5 = 0.75
    \]
    However, if $i_1$ and $i_2$ are interactions of two different users with the service $y_1$, then the recommendation score from the first user is 
    \[
    \sigma(y_1,\{i_1\}) = \mu(\alpha_1|_{\{i_1\}}){\times}0.5 + \mu(\alpha_2|_{\{i_1\}}){\times}0.5 = 0{\times}0.5 + 1{\times}0.5 = 0.5
    \]
    Similarly, the recommendation score from the second user is 
    \[
    \sigma(y_1,\{i_2\}) = 1
    \]
    \item If another trust provider uses the fresh-biased evidence selection then
        \[
    \sigma'(y_1) \simeq \mu(\alpha_1){\times}0.33 + \mu(\alpha_2){\times}0.67 = 0.835
    \]
    and
        \[
    \sigma'(y_1,\{i_1\})\simeq \mu(\alpha_1|_{\{i_1\}}){\times}0.33 + \mu(\alpha_2|_{\{i_1\}}){\times}0.67 = 0.67~.
    \]
    In other words, this trust provider is more likely to recommend the service $y_1$ than the previous trust provider.
    \item In this particular example, the values taken by the scoring mechanism $\mu$ is in $[0,1]$. By Lemma~\REF{lem:proba-scoring} below, the recommendation score is also in $[0,1]$.
    \item $\mu$ is $\rho$-monotonic, that is, if $\rho(x)\leq\rho(x')$ then $\mu(\alpha[i\MapTo x])\leq \mu(\alpha[i\MapTo x'])$ for every feedback trace $\alpha$. Here $\alpha[i\MapTo x]$ substitutes $\alpha(i)$ with $x$ or extends $\alpha$ with $\{i\MapTo x\}$ if $i$ is not in the domain of $\alpha$.
\end{itemize}
\end{example}

\subsection{Soundness of trust scoring}
The following lemma is a natural consequence of the fact that we choose feedback probabilistically for a given interaction. It implies two important theorems which we elaborate in the sequel.

\begin{lemma}
\label{lem:sum1}
For every finite set of interactions $A$, we have
\begin{equation}
\label{eqn:sum1}
\sum_{\substack{\EvidenceMap\Fcomp\alpha=\mathrm{id}\\ \alpha^{-1}(X)=A}} \prod_{\substack{\alpha(i)=x}}\omega(i,x) = 1
\end{equation}
\begin{proof}
The proof is by induction on the set $A$. 
%\begin{itemize}
%\item 

$\bullet$ If $A = \emptyset$ then the sum in Equation~\REF{eqn:sum1} is only coming from the empty feedback trace. By convention, the product over an empty set is $1$ so that the sum in Equation~\REF{eqn:sum1} is also $1$.
%\item 

$\bullet$ Let $A\subset I$ be a finite set and $j\in I\Setminus A$. We have 

{\footnotesize
\begin{Reason}
\Step{}{
	\sum_{\substack{\EvidenceMap\Fcomp\beta=\mathrm{id}\\ \beta^{-1}(X)=A\cup\{j\}}} \prod_{\substack{\beta(i)=x}}\omega(i,x)
}
\StepR{$=$}{\tiny{$\beta {=} \alpha{\cup}\{j\MapTo z\}$ for $\alpha{\leq}\beta$ and $z{\in}\EvidenceMap^{-1}(j)$}}{
	\sum_{\substack{\EvidenceMap\Fcomp\alpha=\mathrm{id}\\ \alpha^{-1}(X)=A}} \sum_{z\in\EvidenceMap^{-1}(j)}\prod_{\substack{\alpha(i)=x}}\omega(i,x)\omega(j,z)
}
\StepR{$=$}{\tiny{Continuity of mult. by $\prod_{\alpha(i)=x}\omega(i,x)$}}{
	\sum_{\substack{\EvidenceMap\Fcomp\alpha=\mathrm{id}\\ \alpha^{-1}(X)=A}} \prod_{\substack{\alpha(i)=x}}\omega(i,x)\left(\sum_{z\in\EvidenceMap^{-1}(j)}\omega(j,z)\right)
}
\StepR{$=$}{\tiny{Equation~\REF{eqn1}}}{
	\sum_{\substack{\EvidenceMap\Fcomp\alpha=\mathrm{id}\\ \alpha^{-1}(X)=A}} \prod_{\substack{\alpha(i)=x}}\omega(i,x)
}
\StepR{$=$}{\tiny{Induction Hypothesis}} { 1 }
\end{Reason}
}
%\footnotesize
%\end{itemize}
Hence, Equation~\REF{eqn:sum1} always holds.
\end{proof}
\end{lemma}

The following theorem shows that we can safely restrict to using scores between $[0,1]$ as in  most traditional trust formalisation.

\begin{theorem}\label{lem:proba-scoring}
If $\mu(\alpha)\leq 1$ for every feedback trace $\alpha$, then $\sigma(A)\leq 1$ for every set of interaction $A$ (when it is defined).
\begin{proof}
Let $A\subseteq I$ be a finite set of interactions. The equation $\sigma(A)\leq 1$ simply follows from Lemma~\ref{lem:sum1} as follows.
\begin{Reason}
\Step{}{
	\sigma(A)
}
\StepR{$=$}{\footnotesize{Definition~\ref{scoring}}}{
	\sum_{\substack{\EvidenceMap\Fcomp\alpha=\mathrm{id}\\ \alpha^{-1}(X)=A}} \mu(\alpha)\prod_{\substack{\alpha(i)=x}}\omega(i,x)
}
\StepR{$\leq$}{\footnotesize{$\mu(\alpha)\leq 1$}}{
	\sum_{\substack{\EvidenceMap\Fcomp\alpha=\mathrm{id}\\ \alpha^{-1}(X)=A}} \prod_{\substack{\alpha(i)=x}}\omega(i,x)
}
\StepR{$=$}{\footnotesize{Lemma~\ref{lem:sum1}}} { 1 }
\end{Reason}
The infinite case follows from the fact that the limit of a convergent sequence of real numbers in $[0,1]$ is in that interval (compactness of the unit interval). 
\end{proof}
\end{theorem}

Note that, even if $\mu(\alpha)\leq 1$ for every feedback trace $\alpha$ but $A$ is infinite, then $\sigma(A)$ does not necessarily exist. For instance, let us take a system where there is exactly one feedback for each review and the feedback ratings are alternating back and forth from $0$ to $1$, for each successive interaction. If $\mu$ simply returns the latest rating in a feedback trace then it will also take alternating values. In this case, we obtain a  sequence of set of interactions such that $\sigma(A_n) = 0$ (resp. $1$) and $\sigma(A_{n+1}) = 1$ (resp. $0$). This sequence does not have a limit.

The following theorem is one of our main theoretical result ensuring the soundness of the recommendation score calculation even at the limits.

\begin{theorem}
\label{thm:convergence} 
Any recommendation trust scoring $\sigma$ is well-defined and $\sigma(A)\leq|A|$ for every finite set $A\subseteq I$. That is, the right hand side of Equation~\REF{eqn:scoring} always converges as long as the context $A$ is finite.
\begin{proof}
Firstly, it follows from Lemma~\ref{lem:sum1} that $\sigma(A)\leq|A|$. This means that the right hand side of Equation~\REF{eqn:scoring} is a bounded series of real numbers with non-negative terms. The general construction of this series is shown in Figure~\ref{fig:infinite-feedback}. We conclude that the series is convergent.
\end{proof}
\end{theorem}

\begin{figure}
\begin{center}

\tikzset{every picture/.style={line width=0.75pt}} %set default line width to 0.75pt        

\begin{tikzpicture}[x=0.75pt,y=0.75pt,yscale=-1,xscale=1]
%uncomment if require: \path (0,266); %set diagram left start at 0, and has height of 266

%Straight Lines [id:da0999274269864664] 
%\draw    (41,70.25) -- (41,207.75) ;
%\draw [shift={(41,209.75)}, rotate = 270] [color={rgb, 255:red, 0; green, 0; blue, 0 }  ][line width=0.75]    (10.93,-3.29) .. controls (6.95,-1.4) and (3.31,-0.3) .. (0,0) .. controls (3.31,0.3) and (6.95,1.4) .. (10.93,3.29)   ;
%Straight Lines [id:da6818114636215183] 
\draw    (260,70.25) -- (260,207.75) ;
\draw [shift={(260,209.75)}, rotate = 270] [color={rgb, 255:red, 0; green, 0; blue, 0 }  ][line width=0.75]    (10.93,-3.29) .. controls (6.95,-1.4) and (3.31,-0.3) .. (0,0) .. controls (3.31,0.3) and (6.95,1.4) .. (10.93,3.29)   ;
%Shape: Rectangle [id:dp26553595828175924] 
\draw  [dash pattern={on 4.5pt off 4.5pt}] (61,10) -- (172,10) -- (172,30) -- (61,30) -- cycle ;

% Text Node
\draw (59,72) node [anchor=north west][inner sep=0.75pt]   [align=left] {$\displaystyle x_{11}$};
% Text Node
\draw (59,101) node [anchor=north west][inner sep=0.75pt]   [align=left] {$\displaystyle x_{12}$};
% Text Node
\draw (89,72) node [anchor=north west][inner sep=0.75pt]   [align=left] {$\displaystyle x_{21}$};
% Text Node
\draw (89,101) node [anchor=north west][inner sep=0.75pt]   [align=left] {$\displaystyle x_{22}$};
% Text Node
\draw (59,161) node [anchor=north west][inner sep=0.75pt]   [align=left] {$\displaystyle x_{1n}$};
% Text Node
\draw (88.5,161) node [anchor=north west][inner sep=0.75pt]   [align=left] {$\displaystyle x_{2n}$};
% Text Node
\draw (147.5,72) node [anchor=north west][inner sep=0.75pt]   [align=left] {$\displaystyle x_{m1}$};
% Text Node
\draw (147.5,101) node [anchor=north west][inner sep=0.75pt]   [align=left] {$\displaystyle x_{m2}$};
% Text Node
\draw (147,161) node [anchor=north west][inner sep=0.75pt]   [align=left] {$\displaystyle x_{mn}$};
% Text Node
\draw (220.5,72) node [anchor=north west][inner sep=0.75pt]   [align=left] {$\displaystyle \alpha _{1}$};
% Text Node
\draw (220.5,101) node [anchor=north west][inner sep=0.75pt]   [align=left] {$\displaystyle \alpha _{2}$};
% Text Node
\draw (220.5,161) node [anchor=north west][inner sep=0.75pt]   [align=left] {$\displaystyle \alpha _{n}$};
% Text Node
\draw (64,12) node [anchor=north west][inner sep=0.75pt]   [align=left] {$\displaystyle i_{1}$};
% Text Node
\draw (94,12) node [anchor=north west][inner sep=0.75pt]   [align=left] {$\displaystyle i_{2}$};
% Text Node
\draw (152,12) node [anchor=north west][inner sep=0.75pt]   [align=left] {$\displaystyle i_{m}$};
% Text Node
%\draw (20,206) node [anchor=north west][inner sep=0.75pt]  [rotate=-270] [align=center] {Stream of feedback};
% Text Node
\draw (264,192) node [anchor=north west][inner sep=0.75pt]  [rotate=-270] [align=center] {Sequence of \\feedback traces};
% Text Node
\draw (118,72) node [anchor=north west][inner sep=0.75pt]   [align=left] {$\displaystyle \cdots $};
% Text Node
\draw (118,101) node [anchor=north west][inner sep=0.75pt]   [align=left] {$\displaystyle \cdots $};
% Text Node
\draw (118,161) node [anchor=north west][inner sep=0.75pt]   [align=left] {$\displaystyle \cdots $};
% Text Node
\draw (75,125) node [anchor=north west][inner sep=0.75pt]  [rotate=-90] [align=left] {$\displaystyle \cdots $};
% Text Node
\draw (105,125) node [anchor=north west][inner sep=0.75pt]  [rotate=-90] [align=left] {$\displaystyle \cdots $};
% Text Node
\draw (165,125) node [anchor=north west][inner sep=0.75pt]  [rotate=-90] [align=left] {$\displaystyle \cdots $};
% Text Node
\draw (75,185) node [anchor=north west][inner sep=0.75pt]  [rotate=-90] [align=left] {$\displaystyle \cdots $};
% Text Node
\draw (234,125) node [anchor=north west][inner sep=0.75pt]  [rotate=-90] [align=left] {$\displaystyle \cdots $};
% Text Node
\draw (105,185) node [anchor=north west][inner sep=0.75pt]  [rotate=-90] [align=left] {$\displaystyle \cdots $};
% Text Node
\draw (165,185) node [anchor=north west][inner sep=0.75pt]  [rotate=-90] [align=left] {$\displaystyle \cdots $};
% Text Node
\draw (234,185) node [anchor=north west][inner sep=0.75pt]  [rotate=-90] [align=left] {$\displaystyle \cdots $};
% Text Node
\draw (118,15) node [anchor=north west][inner sep=0.75pt]   [align=left] {$\displaystyle \cdots $};
% Text Node
\draw (44,-8) node [anchor=north west][inner sep=0.75pt]   [align=center] {Finite set $A$ of interactions};
% Connection
\draw    (70,34) -- (70,66) ;
\draw [shift={(70,68)}, rotate = 270] [color={rgb, 255:red, 0; green, 0; blue, 0 }  ][line width=0.75]    (10.93,-3.29) .. controls (6.95,-1.4) and (3.31,-0.3) .. (0,0) .. controls (3.31,0.3) and (6.95,1.4) .. (10.93,3.29)   ;
% Connection
\draw    (100,34) -- (100,66) ;
\draw [shift={(100,68)}, rotate = 270] [color={rgb, 255:red, 0; green, 0; blue, 0 }  ][line width=0.75]    (10.93,-3.29) .. controls (6.95,-1.4) and (3.31,-0.3) .. (0,0) .. controls (3.31,0.3) and (6.95,1.4) .. (10.93,3.29)   ;
% Connection
\draw    (160,34) -- (160,66) ;
\draw [shift={(160,68)}, rotate = 270] [color={rgb, 255:red, 0; green, 0; blue, 0 }  ][line width=0.75]    (10.93,-3.29) .. controls (6.95,-1.4) and (3.31,-0.3) .. (0,0) .. controls (3.31,0.3) and (6.95,1.4) .. (10.93,3.29)   ;
% Connection
\draw    (217,77) -- (179,77) ;
\draw [shift={(177,77)}, rotate = 360] [color={rgb, 255:red, 0; green, 0; blue, 0 }  ][line width=0.75]    (10.93,-3.29) .. controls (6.95,-1.4) and (3.31,-0.3) .. (0,0) .. controls (3.31,0.3) and (6.95,1.4) .. (10.93,3.29)   ;
% Connection
\draw    (217,106) -- (179,106) ;
\draw [shift={(177,106)}, rotate = 360] [color={rgb, 255:red, 0; green, 0; blue, 0 }  ][line width=0.75]    (10.93,-3.29) .. controls (6.95,-1.4) and (3.31,-0.3) .. (0,0) .. controls (3.31,0.3) and (6.95,1.4) .. (10.93,3.29)   ;
% Connection
\draw    (217,166) -- (179,166) ;
\draw [shift={(178,166)}, rotate = 360] [color={rgb, 255:red, 0; green, 0; blue, 0 }  ][line width=0.75]    (10.93,-3.29) .. controls (6.95,-1.4) and (3.31,-0.3) .. (0,0) .. controls (3.31,0.3) and (6.95,1.4) .. (10.93,3.29)   ;

\end{tikzpicture}
\[
\sigma(A) = \sum_{l=1}^{\infty}\mu(\alpha_l)\prod_{k=1}^{m}\omega(i_k,x_{kl})
\]

\begin{quote}
In this diagram, each interaction $i_k$ has an infinite stream of feedback and each feedback trace is constructed by selecting a review from each interaction using the weight function $\omega$. This picture shows the combination of selecting $m$ feedback for each interaction. That is, the infinite matrix $x_{kl}$ is built by enumerating elements of a set similar to $\mathbb{N}^m$. We then obtain a sequence of feedback traces $(\alpha_l)_l$.
\end{quote}

\caption{Computation of recommendation for an infinite stream of feedback.}\label{fig:infinite-feedback}
\end{center}
\end{figure}

If $A$ is infinite, then $\sigma(A)$ is not necessarily finite. A simple counterexample is to take the sum of individual review ratings for a scoring mechanism. We can assume that there is only one feedback per interaction and each feedback has a rating of $1$. In this case, we can build an increasing sequence of set of interactions $(A_n)_n$ such that $\sigma(A_n) = n$.

%%%
\section{System architecture and implementation} 
\label{proposed-architecture}
% Not sure which architecture we want to go with but here is my idea for one without a second private blockchain providing attributes.

% \begin{enumerate}
%     \item A user requests or makes payment to a trust provider in exchange for trust values on multiple resource providers 
%     \item That user then decides which resource provider to go with. 
%     \item The user makes a request of access which the resource smart contract will emit a PoAR in response
%     \item The user then access of interacts with the resource, using up the token from 3a and creating a PoI
%     \item The user then provides feedback on the interaction and is rewarded with some token/s
%     \item The trust provider must then maintain the feedback state and recalculate any of the trust scores that have been interacted with since the last update
%     \item TP's then update their smart contract with the new trust values
% \end{enumerate}

%%% SP: I commented this architecture of AH
\iffalse
\begin{figure}
    \centering
    \includegraphics[scale=0.25]{images/trust_architecture.png}
    \caption{New Architecture of the proposed blockchain platform implementation.
    %{\color{blue} TR: this figure needs to be updated }
    %{\color{red}AH: Has been updated, let me know what needs to be changed}
    }
    \label{fig:architecture}
\end{figure}
\fi
%%%

This section shows how we design and implement the formalism of Section~\ref{sec-framework} on a blockchain platform. We use various basic properties of a blockchain (e.g., immutability, transparency, decentralised control, etc.) to ensure that Property~\ref{storage} and~\ref{function} are satisfied. This guarantees that the theoretical framework we have developed in Section~\ref{sec:scoring} is sound and practical.

\subsection{System architecture}
The architecture is illustrated in Figure~\ref{fig:architecture}.  The core components of the architecture are users, user device, trust provides and smart contracts. In our case, users can be both resource providers and consumers. User devices can be seen as the smart mobile device carried by the users. These devices can be used to store information, access a resource, deploy smart contracts, as well as communicating with one another. Trust provides maintain the trust scores. Finally, smart contacts are collection of code and data that are used for executing agreements between two parties, and stored on a blockchain. We use three smart contracts, (1) the \textit{resource smart contract} that handles access to a resource, (2) the \textit{feedback smart contract} that handles the reviews submitted by the end-users, and (3) the \textit{trust provider's smart contract} that helps the trust provides to maintain trust scores. We make use of these various smart contracts to ensure secure access to any restricted resource. 

Note, this architecture is based on our previous works \cite{8894097, 8752021} to handle feedback ratings on the blockchain itself. The architecture presented in \cite{8894097, 8752021} used fully featured attribute-based access control to handle policy management. In this paper, in particular, we remove the attribute-based access control in favour of a generalised access scheme where consumers are able to decide what resource to use based on any metric that is available, be it price, ease of access or location. The system is composed of a public blockchain that keeps track of all delegated access rights, consumer interactions and consumer feedback that is directly linked to one consumer interaction. The correctness and security of such a system is explained in more details in~\cite{8894097}. 

\begin{figure}
    \centering
    \includegraphics[scale=0.6]{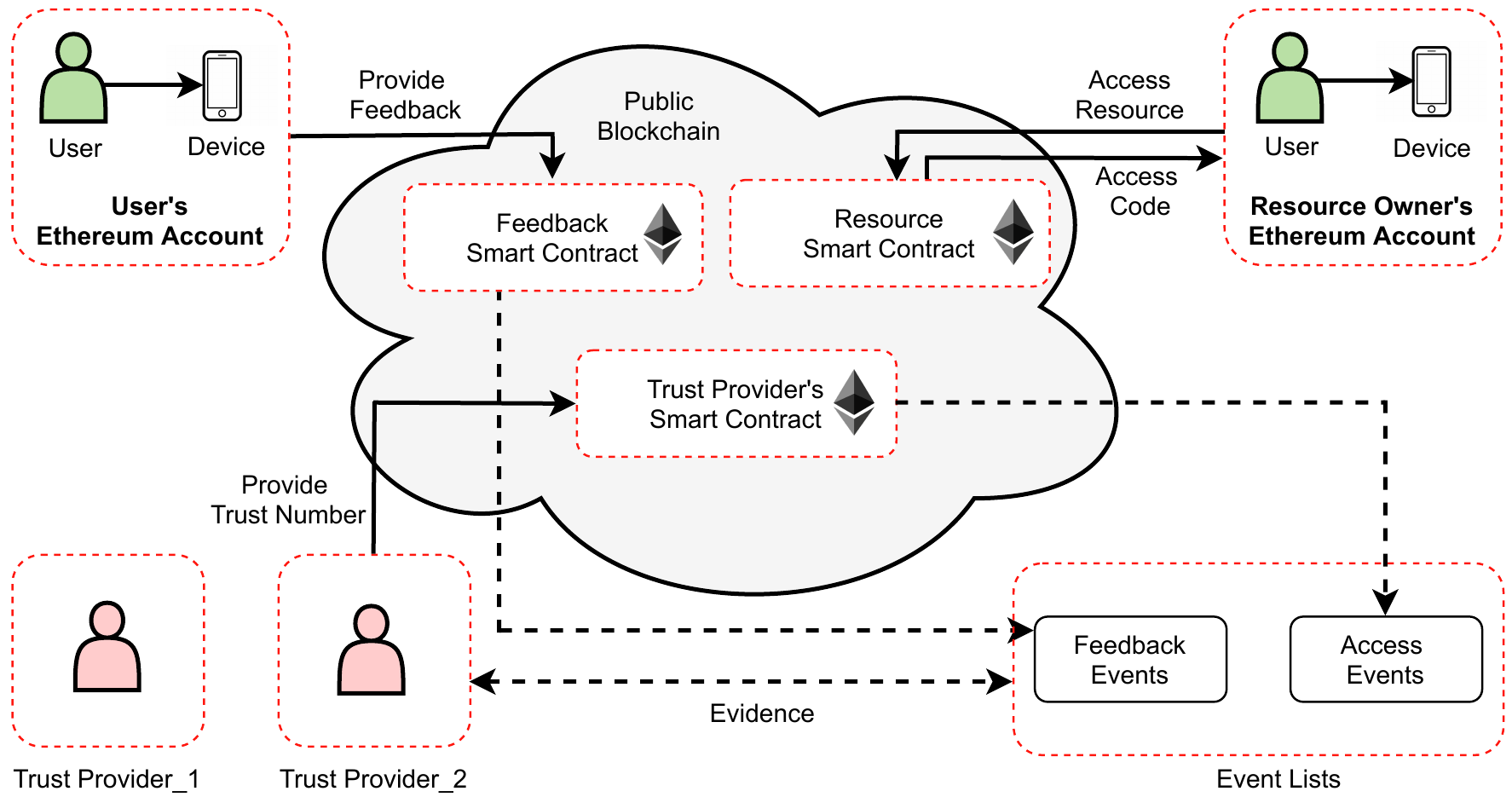}
    \caption{Architecture of the proposed blockchain platform implementation.}
    \label{fig:architecture}
\end{figure}

Note, in this paper, we extend our previous architecture (of \cite{8894097, 8752021}) by adding the following two more crucial components (namely, feedback smart contract and trust providers) to handle reviews and trust score calculations. These new components are only interacting with the public blockchain. A brief description for each of them is given as follows: %They are discussed as follows:

\subsubsection{Feedback smart contract}
This smart contract handles reviews submitted by end-users in the system. In essence, it receives a review rating and ensures that the review is linked to an interaction. To achieve this, the feedback smart contract performs the following actions.
\begin{itemize}
    \item It ensures that a submitted review has the following parts:
    \begin{itemize}
        \item[-] Address of the user account that submits the review, 
        \item[-] Details of the interaction reviewed, and
        \item[-] A review rating.
    \end{itemize}
    \item It checks the existence of the interaction by locating the relevant event in the blockchain.
    \item It checks that the user account that submits the review is the same account that interacted with the service.
    \item It sanitises the rating and if all of the above checks pass, then an event is emitted which contains the review details. In this case, the review becomes a feedback.
\end{itemize}

An event emitted by the Feedback smart contract is what we would call an \emph{feedback}. Since they are provably linked to an interaction, the pair of interaction and feedback events constitutes an \emph{evidence}. This evidence is the backbone of the trust scoring by the trust providers.

The proposed architecture and the implementation of the feedback smart contract guarantee Property~\ref{storage} and~\ref{function}. In particular, there is indeed a well-defined evidence map $\EvidenceMap$ which is built iteratively as the set of evidence grows. Moreover, the correctness of the system can be verified by all parties interacting with the public blockchain.

\subsubsection{Trust providers} 
These entities are responsible for implementing the trust scoring functions and making the output available to the end-users, possibly in exchange for some access fee. 

Since the soundness of the theoretical framework is ensured mainly by the feedback smart contract, a trust provider complements that by choosing a scoring mechanism $\mu$ and an evidence selection $\omega$ to implement. Since we allow the existence of multiple trust providers, the choice of these functions can vary from one implementation to another. The important requirement is that the conditions in Definition~\ref{def:feedback-selection} and~\ref{def:scoring-mechanism} are satisfied and that all trust providers are using the exact same set of evidences stored on the blockchain.

Similarly, to the choice of $\mu$ and $\omega$, we leave it to the trust providers to use an adequate implementation. In our prototype, we have implemented a cached trust scoring (i.e., scores are cached on the blockchain itself and are re-computed periodically) with $\mu$ computing an average rating over the past interactions and $\omega$ is a deterministic evidence selection.

%%%
\subsection{Communication}
\label{design}
\begin{figure}
    \centering
    \includegraphics[scale=.65]{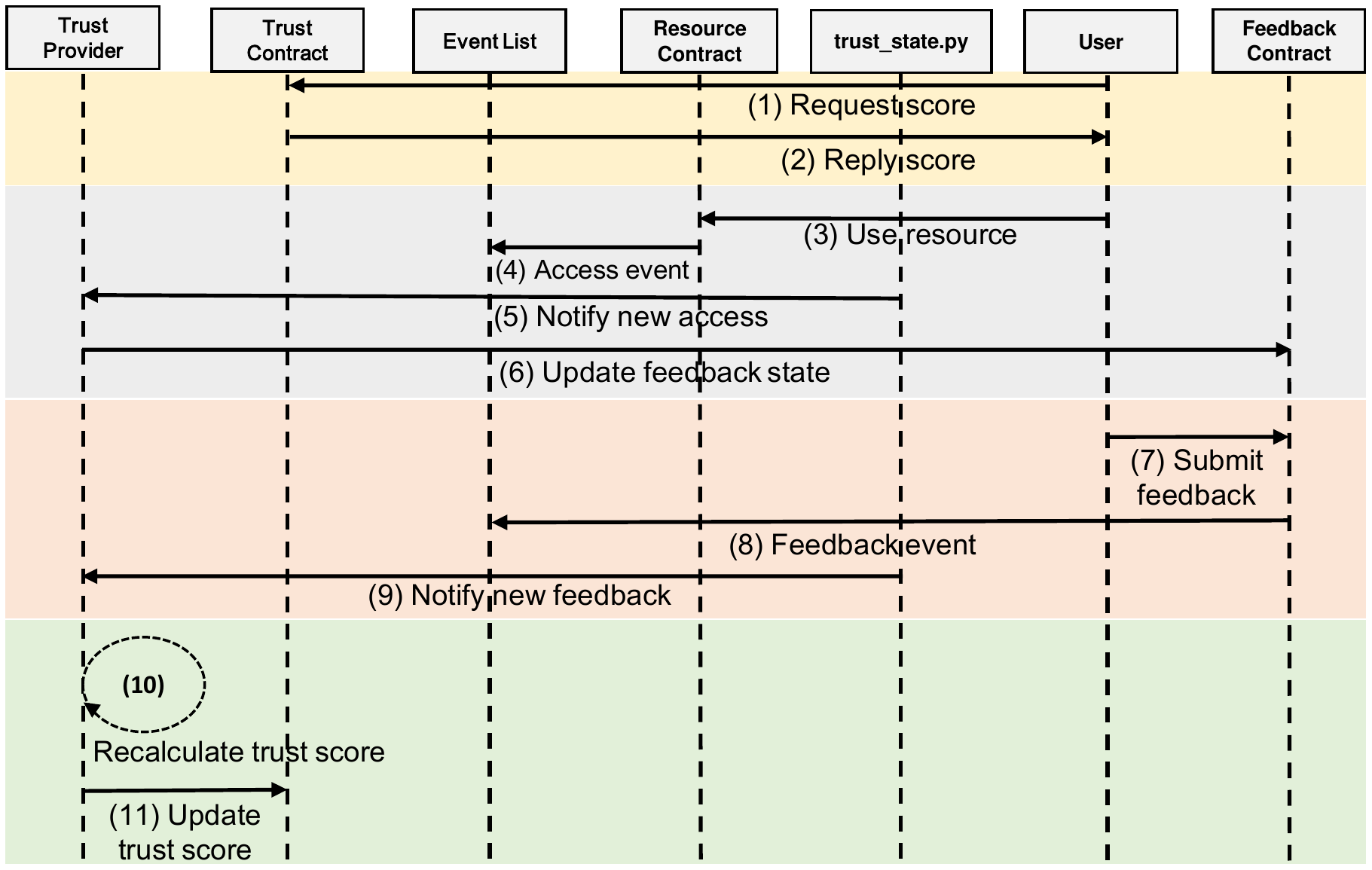}
    \caption{Communication between the various components of the system. Note, for simplicity, a smart contract is written as contract.}
    \label{fig:protocol}
\end{figure}

An overview of the communication between the various components of the system is depicted in Figure~\ref{fig:protocol}. %As in the previous section, the communication protocol between different entities in our system is also extended from our previous work~\cite{8894097}. 
We encapsulate the access protocol with the choice resolution using trust scores and the feedback rating submission after an interaction. This communication diagram is a simplified illustration of shown interactions among various components in Figure~\ref{fig:architecture}.

%The extended communication between the various principals, seen from the perspective of a student who wants to use one of the university's printer, progresses as follows:
%{\color{red} (SP, I want to remove student and printer from here to make it general.)

The extended communication between the various principals, seen from the perspective of a user who wants to use a resource, progresses as follows:

%{{\color{red}}AH: Let me know which version of feedback you want, TR: Thanks}
%\begin{figure}
%    \centering
%    \includegraphics[scale=.5]{images/feedback-snippet.png}
%    \label{fig:feedback-event}
%\end{figure}

%%% Here

\begin{itemize}[label={}]
    %\item Step 1: When a student wants to use one of the printers, he has to choose a broker - whether to use the service provided by his department or the student union. This choice is resolved by querying for the respective trust scores from one or many trust providers. This can be another smart contract or even an off-chain service, if required.
    
    \item Step 1: When a user wants to use a resource, they need to choose a broker - whether to use the service provided by an individual (or a group of users). This choice is resolved by querying for the respective trust scores from one or many trust providers. This can be another smart contract or even an off-chain service, if required.

    \item Step 2: The trust provider retrieves a pre-computed score or perform an on-demand trust score calculation. If the request is through a smart contract, the output of this interaction is pushed as an encrypted event using the user's Ethereum account's public key. This ensures that, if the trust provider charges a fee for its service, then only the requesting user will be able to read this output off the blockchain.
    \item Step 3: Once a broker is chosen, the user will request to use the resource. Whether the user will be required to pay for this use or some other access control scheme is used is dependent on the resource provider. 
    
    % \begin{itemize}
    %     \item[-] Attributes, residing inside of the private blockchain, are queried by the delegation smart and access control conditions are checked.
    %     \item[-] If the queried attributes satisfy the access control conditions then the delegation smart contract emits an event which now serves as the proof of access right.
    % \end{itemize}
    
    \item Step 4: Once the resource has been used an event will be generated on the blockchain which will be the the \emph{proof of resource use}. In Figure~\ref{fig:access-event}, we show the content of a resource access event.
    
\begin{figure}[ht]
\begin{center}
\begin{small}
\begin{verbatim}
    // Event emitted once a resource is used
    event printer_access (
        address user,
        address resource,
        uint uid          );
\end{verbatim}
\end{small}
\caption{The content of a resource access event.} \label{fig:access-event}
\end{center}
\end{figure}
    % Once the student obtains a proof of access right, he can now access the printer (directly or remotely). The printer (or the resource manager) will check for the existence of the proof of access right event on the blockchain and verifies the access conditions. If the check succeeds, the job is printed and a proof of interaction event is emitted.
    
    \item Step 5: The maintainers of the feedback state i.e., trust providers, will be notified of the new resource access since the event is broadcasted on the public chain. 
    % Once the student successfully prints something, he can be incentivised to submit a review. The review must be submitted with the same account attached to the proof of interaction, otherwise it won't become a feedback. This is handled by the feedback smart contract.
    
    \item Step 6: The trust providers will then update the feedback smart contract on the blockchain to update the feedback state. This transaction must be completed by a registered trust provider and ensures that feedback submitted is attached to a use of a resource. 
    % The feedback smart contract, with the help of a background service if required, checks the validity of the review according to the previous section. If the review is valid then it becomes a feedback. A snippet of a feedback event is given in Figure~\ref{fig:feedback-event}.
    
    \item Step 7: If the user is willing and/or incentivised they can then leave feedback for that resource. This feedback is linked to the access event via the unique id and the user's ethereum account address. 
    
    \item Step 8: The feedback smart contract will validate the feedback using the users address and submitted id. If valid the smart contract will emit a feedback event with the users review. In Figure~\ref{fig:feedback-event}, we show the content of a feedback event.
    
    \begin{figure}[ht]
    \begin{center}
    \begin{small}
    \begin{verbatim}
        // Event emitted once feedback has been submitted and validated
        event feedback (
            address submitter,
            address delegator,
            uint rating,
            uint uid
        );
    \end{verbatim}
    \end{small}
    \caption{The content of a feedback event.}\label{fig:feedback-event}
\end{center}
\end{figure}

    \item Step 9: Similar to step 5, the trust providers monitoring the events from the blockchain will be notified that there is a new feedback event. 
    
    \item Step 10: Trust providers can then add this verified feedback into their own trust calculations. Whether this is done in real time, batches or on demand is dependent on the individual trust providers. 
    
    \item Step 11: Once the new trust score(s) are calculated the trust providers can update the scores on the smart contract ready for the next user to request them. 
    % All trust providers now have an extended set of evidence to update their trust calculations.
\end{itemize}

It is clear that the system is self-sustaining on the blockchain which we believe is another advantage in addition to the benefits we listed in Section~\ref{sec:motivation}.

%%%
\section{System evaluation}
\label{evaluation-results}

This section outlines two key evaluations of our system. Firstly, a discussion on how our system guarantees the two properties of our solution. Secondly, we will discuss the results observed when the test implementation was evaluated. 
The experiments we have conducted, demonstrate the feasibility and performance of our system. %{\color{red} THIS MAY GO AWAY -- There are no experiments comparing with other systems as, to the best of our knowledge, there is no properly comparable proposal.} 

\subsection{Guarantee of the fundamental properties}\label{sec:evaluation6.1}
Here we recall the two properties (discussed in Section~\ref{sec-framework}) for convenience:

\begin{property}{1}\label{storage}
Evidence are accessible to all parties interacting with the system. This means that the sets $X$, $I$ and $Y$ are stored on a shared medium --- the blockchain.
\end{property}
\begin{property}{2}\label{function}
A feedback is always supported by an interaction. 
\end{property}

Property 1 is guaranteed through the use of a blockchain. The key features of using a blockchain are, immutability, transparency and decentralised control. This implementation uses a private Ethereum blockchain, forcing the users of the system to identify themselves for access to be allowed. Although there are privacy concerns with the use of a private blockchain, this solution does provide all users with direct access to the access requests, resource interactions and feedback submissions. When any of these actions occur, an event is generated and stored on the blockchain. This creates an immutable digital history of actions performed in the system that any user can verify, fulfilling property 1.

Property 2 is also guaranteed but is not as trivial to prove. The implementation has a component labelled ``feedback state'' which is owned, maintained and deployed by a trust provider or consortium of trust providers. This component is responsible for monitoring the actions occurring on the blockchain and maintaining a world state that records what users have had a valid interaction with a resource. This world state is what is used by the feedback smart contract to validate any feedback submission. Each time a user submits a review, the world state is checked to see if that user has had an interaction. If they have, the review is transformed into feedback evidence and stored on the blockchain, if not, the review is rejected and the transaction is reverted. Using smart contracts and a registering/de-registering process, trust providers will ensure that each feedback is linked to an interaction. For a user with $m$ interactions, they can leave $n$ feedbacks, all of which are linked to a valid interaction. However, using a deterministic evidence selection, only one feedback per interaction will be used to calculate trust scores.

This way, the blockchain will only ever store reviews that has a supporting interaction, hence, guaranteeing property 2.

\subsection{Testing environment}
To demonstrate the feasibility of our proposed architecture, we have implemented the prototype using a private Ethereum blockchain. Our prototype is built on a Dell Latitude 7490 Notebook with 8 GB of memory and an Intel Core i5 processor. We used the go-ethereum implementation Geth, version 1.10.2, to generate and interact with the blockchain networks. Python version 3.6, with the Web3, version 5.18, library to generate and execute transactions. The smart contracts are written using Solidity, version 0.9, which is a high-level language for implementing blockchain smart contracts. 

The testing environment mimicked the system architecture with less nodes operating. The private test network consists of three nodes: a miner and two peers. One peer acts as the resource owner and trust providers, while the other acts as the user requesting trust scores, using the resource and submitting feedback.

The trust providers are also responsible for maintainingy \textit{trust.py} which is a state maintenence program responsible for updating the feedback state once resource use is detected. Only one \textit{trust.py} program, is executed as there is only one trust provider in the test network. \textit{Trust.py} is vital for guaranteeing property 2 for our system, however for testing purposes it is assumed that each feedback is valid, as to get the maximum throughput as a result.
With that system set up the benchmark rounds were run as follows:
\begin{itemize}
    \item Step 1: Test network is setup and the trust score, feedback and printer resource contract are deployed. 
    \item Step 2: A new user account is created for this round of testing.
    \item Step 3: That user account requests a trust score from the trust contract about the resource. How long it takes to have this trust score returned is recorded. 
    \item Step 4: Simulate accessing the resource 11,110 times so that in the next step each feedback is considered valid.
    \item Step 5: Generate the workload of transactions for this round. The tests were ran with 10, 100, 1000 and 10,000 transactions in the workload to calculate system performance under different workloads. The time taken for all transactions in the workload to be computed by the feedback smart contract and emit a feedback event is recorded. The cost (known as ether) of each transaction is also recorded. 
    \item Step 6: Repeat steps 2-5 as necessary.
\end{itemize}

Running the above testing 10 times, for a total of over 111,000 transactions leads to the following results that is depicted in Figure~\ref{fig:results}. Next, we will discuss these results in terms of performance and feasibility. To calculate average block size and block time we ran a test using a workload of 100,000 transactions.

\subsection{Performance}
We note that from a performance perspective the major challenge is the scalability of the user interaction time, i.e., how long does it take for a user to submit requests to this system and receive confirmation that there request has been served (i.e., steps 1-2 and steps 7-8 of Figure~\ref{fig:protocol}). Below, we discuss the transaction times and costs for, (1) a user requesting trust scores, and (2) a user submitting feedback to the system as well as (3) our systems average block generation time and number of transactions per block. 
%However, before we discuss the interaction time in detail, let us consider the other results e.g., gas cost and throughput in short.

\begin{figure}
    \centering
    \includegraphics[scale=.5]{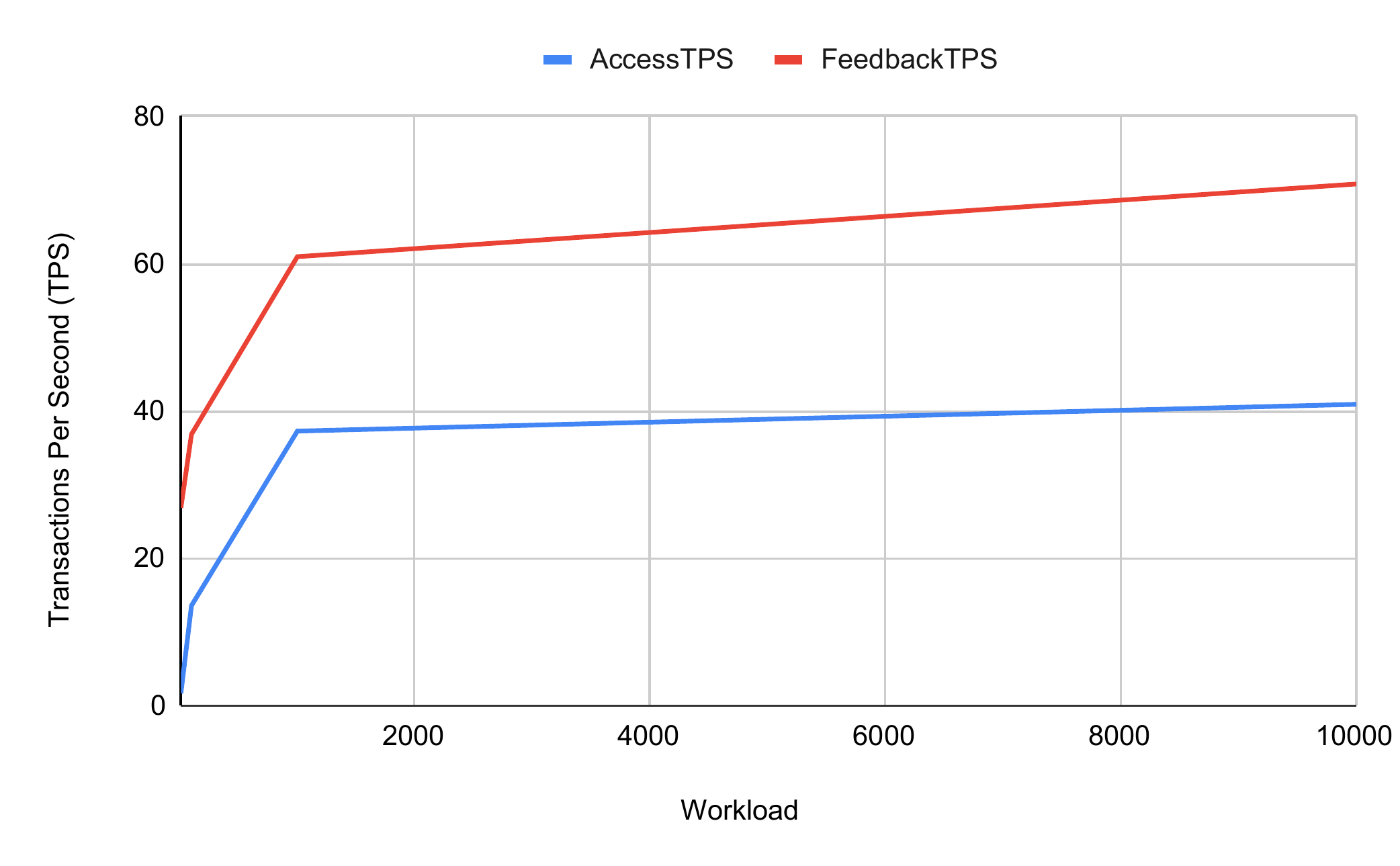}
    \caption{Transaction times for accessing a trust score and submitting feedback}
    \label{fig:accessresults}
\end{figure}

\subsubsection{ Transaction times for accessing trust scores and submitting feedback}

Figure~\ref{fig:accessresults} shows the results from running throughput simulations on our system where a user requesting a trust score from the trust provider and submitting feedback to the system. It can be seen that as work load increases so does the throughput of our system up to about 1000 transactions in the workload. This implies that while the workload is under 100 transactions per second the response time will be near instant.  When the workload is small the throughput is reliant on the block size and block mining time. It can be seen that the transactions per second (tps) improves significantly once the workload increases and saturates around 1500 transactions with 37 tps. Submitting feedback follows the same pattern as accessing a trust score however we see a slightly higher saturation point at just over 60 tps. We believe this is because the complexity of the feedback state is handled off chain by trust providers and so the feedback transaction simply needs to do one check and emit an event, while the access trust score requires the transfer of a specific amount of ether.

% In this case, we consider the case of requesting trust scores from the trust providers. Recall, typically, trust providers calculate their trust scores off-chain. Trust providers can update their trust scores on the world state whenever they want. This value can be requested by any user on the network, since this transaction is a simple exchange of ether for a value, it is a constant time operation that only scales with throughout of the network i.e., workload and mining power. 

\subsubsection{Gas costs for accessing a trust score and submitting feedback}
% which is less than a tenth of a second. We have observed that the average value for this lookup after running 3000 tests is between 6 to 60 ms with an average of 16.3 ms. Our implementation calculates trust in epoch's and updates the final value on the blockchain periodically. Here, we assume that, for a given service, user-agnostic trust score retrieval is free. The time may change if a financial transaction is required, the trust score is fully calculated on-demand or the trust calculation is more complex (e.g., a user-specific recommendation trust score)\footnote{Epoch time and specific trust calculation algorithms are trust provider dependent.}.

\begin{figure}
    \centering
    \includegraphics[scale=.5]{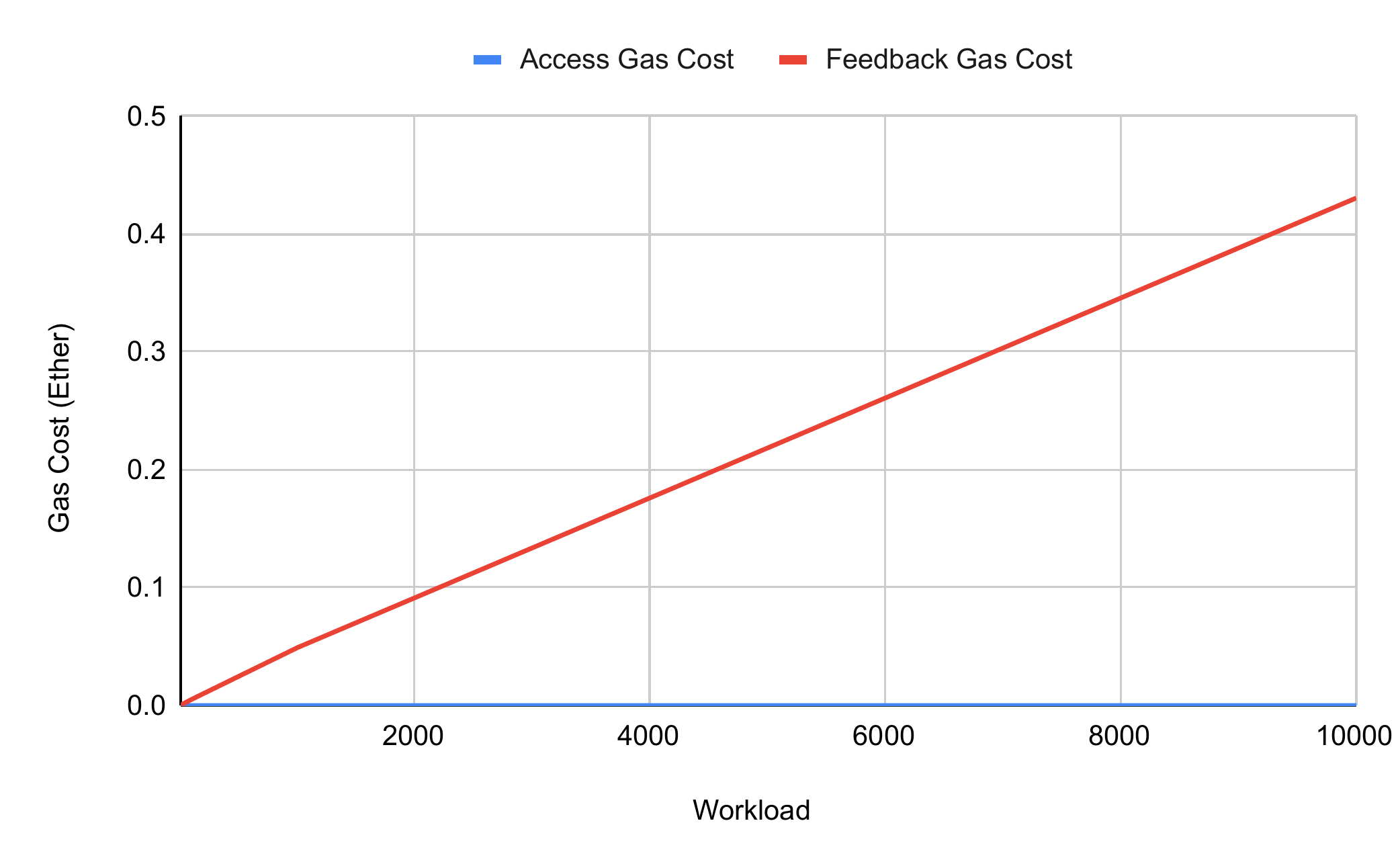}
    \caption{Gas costs for accessing a trust score and submitting feedback}
    \label{fig:results}
\end{figure}

Figure~\ref{fig:results} shows the gas cost in our system for accessing a trust score and submitting feedback. The results indicate that the gas cost increases with workload but the increase is minimal and in a production network would not be noticed by regular users. This allows the gas cost to remain functionally stable, only increasing by a small margin and this is due to dynamic gas pricing by the miner. In other words, as network traffic increases the miner will charge a larger transaction fee resulting in a more expensive transaction for the user. For ease of reading the gas cost for accessing has been multiplied by a scalar of $10^4$.

The gas cost for submitting feedback also increases with workload and will be noticed by users. This is because this transactions reads the state as well, to validate the feedback but then must also emit an event, so as workload increases miners will raise the transaction fee and gas cost which reflects to increase in the cost of the transaction.

\subsubsection{Average block time and number of transactions per block}
\begin{figure}
    \centering
    \includegraphics[scale=.5]{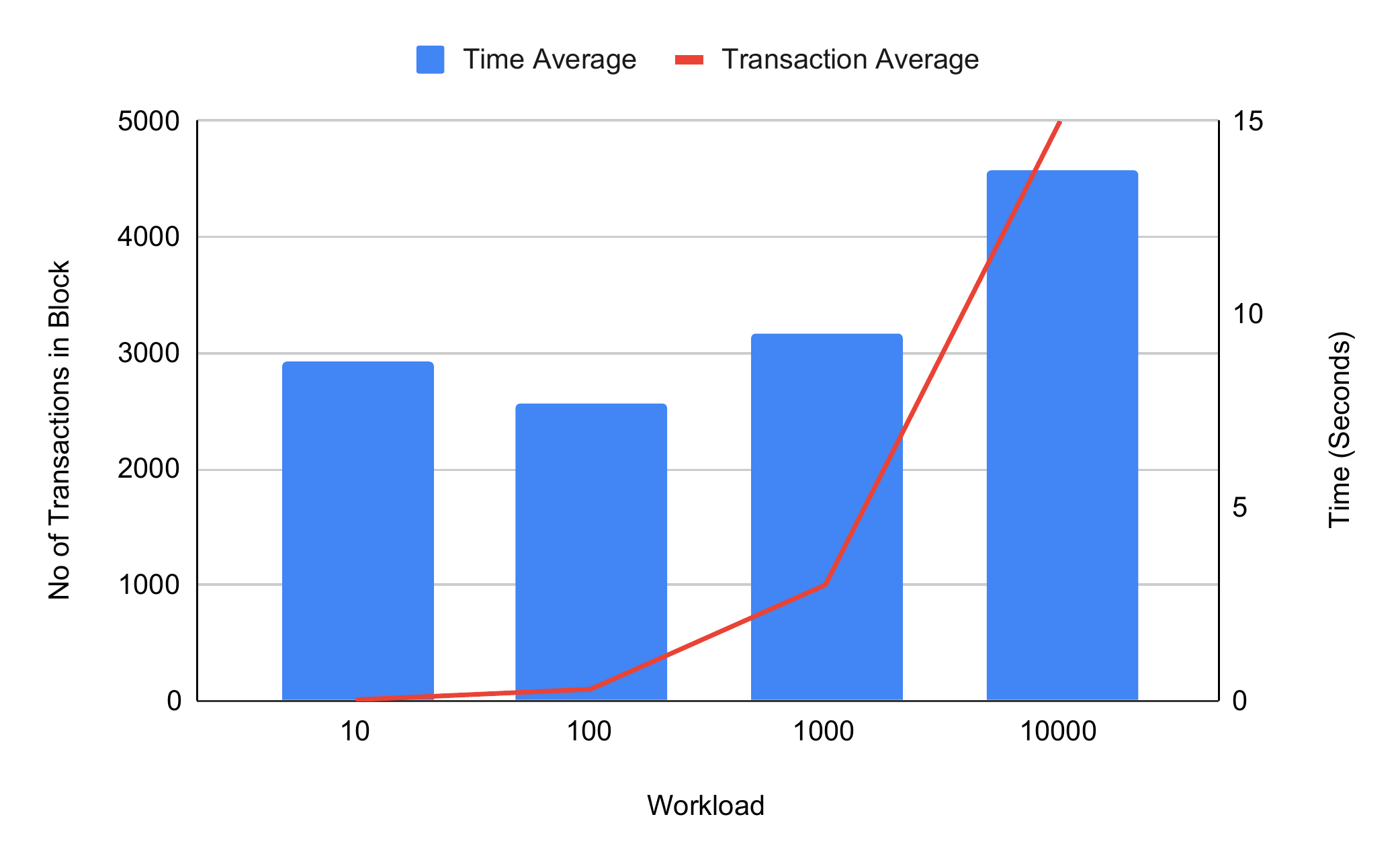}
    \caption{Block generation time and average number of transactions per block}
    \label{fig:blocktimeresults}
\end{figure}

Figure~\ref{fig:blocktimeresults} shows the time to generate blocks and the average number of transactions in a block under different workloads. In order to calculate average block time we simulated the same workloads (10, 100, 1000, 10000) as our results above and recorded how long it took each block to be generated and how many transactions were in that block. Our system makes use of Ethereum's ethash consensus algorithm which will dynamically change the difficulty of the hash puzzle based on current mining power. Ethereum tries to maintain an average block time of 12 seconds and our results follow this pattern. Block size and in turn number of transactions in a block is based on total gas limit which for our tests was set to maximum, allowing up to 5000 transactions into a single block when the workload demands are high. We notice that when the workload is small the number of transactions in a block is equal to the workload i.e., when the workload is 10 the number of transactions in a block is 10, same for 100 and 1000. This is because our feedback and access transactions are designed to be simplistic, requiring less gas and ultimately allowing more transactions into a single block thus increasing our system throughput.

\subsection{Feasibility and comparison}
The achieved results indicate that our system is able to handle large amounts of transactions in a reasonable amount of time. The critical transaction from an end user's perspective is how long it takes to retrieve trust scores and for this purpose our system can be configured to perform this task within a few seconds. The throughput also continues to improve as the workload gets larger indicating that even when the system is performing under stress, accessing the trust scores is still on a scale of seconds. The other key result is how long it takes for feedback to be submitted and accepted. As this function produces an event it is more expensive via gas cost, but this can be alleviated by incorporating a reward system for honest users providing feedback. Lastly, we can compare a part of our solution (i.e., how quickly a feedback/evidence can be verified) to an existing trust based evaluation system \cite{yan2021social}, called social-chain. Note, proposal~\cite{yan2021social} is based on a different consensus algorithm compared to ours. Proposal~\cite{yan2021social} claims that they can handle 1 evidence per second whilst guaranteeing decentralisation, our system currently simulating at around 70 feedback events second with higher workloads. This is a significant improvement and since our system is using Ethereum there is evidence that the system could be easily deployed on a public blockchain e.g., the Ethereum main net. 

For results that show similar trends but for delegated access control, we refer to our previous work in~\cite{8894097}.

% The only transaction that does not scale with the size of the network, in respect to events, is the interaction time. Although this is expected, there are ways around mitigating the impact of this bottleneck process. 

% The first is to limit the number of blocks a user is searching for. For testing purposes, the system was artificially inflated with fake events. This caused a large number of events in a short number of blocks. In reality, a deployed system like this would not expect 10,000 access requests within a couple of minutes, it would realistically be spread out over the day. Given a more realistic spread of events, a resource owner could mandate that requests older than 24 hours are not valid and as a result reduce the search space for the events. 

% The second is to have resource owners maintain a world state, similar to that of trust providers maintaining the feedback state. This would allow resource maintainers to do a simple lookup in the world state to see if the user submitting/requesting the interaction has been granted access. If this were to be implemented it would be expected to see interaction time follow the same trend as feedback time that we achieved in our results.

%%%
\section{Discussion on adversary model and attacks}
\label{sec:attack}
%\subsection{General discussion regarding the results and approach}
%\begin{enumerate}
%    \item Discuss trade offs between where we calculate and store the trust scores i.e on chain, off chain, smart contract
%    \item Discuss choice of trust calculation
%    \item Discuss choice of blockchain platform and architecture?
%\end{enumerate}
%Much of this applies to a centralised trust system, but we’re not centralised in some ways we combine the advantages of both

%\subsection{Adversary Model and Attacks}
In this paper, we assume that the communication occurs over secure channels. Therefore, protocol attacks, e.g., Man-in-the-Middle (MITM), are out of the scope of this paper. There may be different kinds of attacks possible in such a scenario~\cite{GUO20171}. Of more interest to us are attacks specific to trust systems. Attacks on trust systems are generally carried out by hostile actors within the trust system.  These can fall into three categories, (i) trusted entities switching to bad behaviour after a period of good behaviour, (ii) entities making false recommendations, and (iii) a trust provider giving a false response to an enquiry by a user~\cite{4068036}.  Most previous work focuses solely on the first two and not the latter, typically because such systems feature only a single trust provider and trust in it becomes a basic assumption. Our system does provide some protection against a hostile trust provider, but before considering that issue we will examine some of the more typical attacks~\cite{cloudarmor}~\cite{LI2020841}~\cite{SI20191028}~\cite{jayasinghe2018trust}. These include:

\begin{itemize}
    \item Opportunistic service attacks: A malicious node provide good services to increase the trust value held for it. Then it performs malicious activities.
    \item On-off attacks: This can also be seen as a random attack. In this attack, a malicious node performs good and bad activities randomly to avoid being labelled as a low trust node.
    \item Collusion attacks: Multiple hostile actors collude to give misleading feedback.
    \item Sybil attacks: A single hostile actor creates several accounts exploiting multiple identities to provide misleading feedback.
\end{itemize}

Opportunistic service attacks and on-off attacks are examples of an entity using an acquired trusted reputation to achieve a position from which to carry out hostile activities. Collusion and sybil attacks both may make use of good-mouthing (where the feedback given intentionally over-rates the value of an interaction) or bad-mouthing (where the feedback given intentionally under-rates the value of an interaction). Collusion attacks are where a number of system entities launch good-mouthing or bad-mouthing attacks in concert. A sybil attack is similar, but is where a single system entity creates multiple clones of itself to launch good-mouthing or bad-mouthing attacks. 

Some basic features of our proposal should be noted in considering how our system inhibits various forms of attacks. The first of these is that trust providers will only take feedback into account when that feedback is present in the blockchain and linked to a record of an actual interaction. This is unlike  many other systems, where a recommendation can be given even if no actual interaction has taken place. Trust providers in our system can be certain that an interaction has occurred. Second, in the applications that we envisage for our system, payment is required. This means that hostile entities have to expend financial resources to make false recommendations. The recommendation itself is free, but can only be made after a non-zero cost interaction. In most systems, there is no resource cost to making a recommendation.  Finally, feedbacks are time-stamped, allowing trust providers to determine when, and with what frequency, feedbacks are given and whether there are any unusual patterns (e.g., spikes) in feedback provision.

These features make it more difficult for hostile entities in our systems to conduct good-mouthing and bad-mouthing attacks. In the case of bad-mouthing, the hostile entity would have to interact with, and pay for a service from, the entity it wished to bad-mouth. In many other trust systems, there is no way of ensuring that an entity providing a low recommendation has actually interacted with the service provider. This is not as effective a deterrent to good-mouthing, as the recommender and service provider may likely be in collusion and can arrange for a return of funds, but having to arrange for reimbursement adds to the complexity of the attack. In the case of both good and bad-mouthing, these attacks often occur in a short span of time, attempting to overwhelm the genuine feedbacks. The time-stamping of recommendation reports on the blockchain allows these unusual spikes to be detected and discounted.  The number of feedback reports made by a particular recommender can also be seen.  If a recommender makes a series of low feedback reports about a service provider, the trust provider could ignore such reports on the basis that a genuine user would not continue to employ a poor service provider and such reports are likely an instance of bad-mouthing.

Bad-mouthing and good-mouthing from a single entity are unlikely to have much effect on the trust value returned by a trust provider for a service provider.  A successful instance of such attacks requires multiple recommenders if the service provider's reported trust value is to be affected. Hence, the employment of sybil and collusion attacks. In the case of a sybil attack, it is likely that the hostile (cloned) respondents would only interact with the target entity. This can be detected by the lack of presence on the blockchain of any other recommendations from these entities.  Even if the cloned entities do make other recommendations, these will likely be within a very short space of time, as the maintenance of these identities, in an active state, by the initial hostile actor, is unlikely. This especially true if all recommendations made by an entity in its short active lifetime are negative.  While collusion attacks may involve entities with a longer active lifetime than sybil attacks, they again tend to occur in short time periods.  An unusually large burst of negative feedbacks is easily detectable by the trust providers in our system.

An opportunistic service attack is very hard for any trust system to detect if it is aimed at a single user.  However, if all recent feedback for a particular service, from all recommending entities, including those with extensive lifetimes of active behaviour, are negative then the trust provider is able to conclude that the service provider has turned hostile.  Note that in many distributed trust systems, a single poor interaction, even if it was the most recent one, would have limited effect on a recommendation. In our system, due to the recommendations being time-stamped, the pattern of behaviour of service providers across multiple users is visible. By looking at the number of users reporting, the trust provider can make a decision about whether this is a collusion attack or an opportunistic service attack. On-off attacks are likewise hard to detect for most trust systems and our system will also struggle to detect one.  At best, the time-stamping of feedback may allow the trust providers to detect a pattern of switching between good and bad behaviour.

Much of the above discussion would apply to centralised trust systems (i.e., ones where individual interaction reports are made to a central trust manager, which can be queried for recommendations)~\cite{QURESHI2021107647}.  While centralised trust systems, in general, cannot rely on a genuine interaction having occurred, or attackers needing to expend funds to access a service before making a recommendation, centralised trust systems are aware of the time at which a report of an interaction was made. This is, as noted above, very useful in detecting certain forms of attack. Our system delivers this without the drawbacks (single point of failure, etc.) of a centralised system, while guaranteeing that an interaction has occurred.  More than this, our system allows for the existence of multiple trust providers, which users can choose between. Users in our system can audit trust provider performance, as the users also have access to the individual trust reports, something not true in most centralised trust systems. In summary, our system is both decentralised and provides transparent data to both the users and the trust providers to enable them to come to decisions.

This extra availability of information helps protect the system against hostile trust providers, something most previous work does not address.  As the users are able to access the trust reports directly from the blockchain they can evaluate the trust provider responses if they wish. In most trust systems that rely on a third-party trust manager, users have very little, if any, opportunity to verify the veracity of recommendation values. Users in our system can also avail themselves of multiple trust providers to determine the likely veracity of responses. Therefore, our system, both by providing increased access to information and the potential for multiple trust providers, gives some protection against hostile trust providers, an issue not dealt with by other systems~\cite{WU2021108004}.

Note that our system allows not only the trust providers but the general system users access to the recommendations. This potentially allows for new and interesting interactions between the users and the trust providers, e.g., users specifying which feedback trust providers are to use in calculating reported trust values. This could be on the basis of the identity of the entities providing the feedback, the time period in which the feedback was made, or the exact service being provided. While most distributed  trust systems also allow such choice, they require the users to do the calculation. Here the calculations are carried out by the dedicated trust providers.

\section{Related work}
\label{sec-related work}

 Trust is of interest to many areas, including physiology, sociology and computer science, just to name a few.  There is no universally accepted definition for trust~\cite{arai2009defining} and the interpretation can range from philosophical semantics to technical implementation, and everything in between. For instance, trust can be referred as to the honesty, truthfulness or even the reliability of a trustee. It always varies depending on the context and the targeted use. Trust and reputation in information and communication technologies are prime factors for successful communication between two entities. The notion of computational trust dates back to the early nineties. Since then, trust has been widely discussed, formalised and quantified~\cite{Cho-2815595}~\cite{MOUSA201549}.

The notion of trust originates from social interactions. From the point of view of the social sciences, it integrates the idea of social influence. This comes from the characteristics and behaviour of an individual and can be measured as the honesty, cooperativeness and their willingness to offer help within a social group. There are wider definitions for describing trust in social sciences. In psychology, trust represents a belief that says that a trusted person will do what is expected~\cite{evans2009psychology}. This defines the aspect of trust from an internal phenomenon (of a person) that helps to maintain a normal relationship between individuals. Rousseau et~al.~\cite{rousseau1998not} discuss trust as \textit{``a psychological state comprising the intention to accept vulnerability based upon positive expectations of the intentions or behaviour of another''}. Other philosophical variants~\cite{deutsch1962cooperation, 101007-3, 101007978-3} of trust exist in the social context but most of those definitions revolve around the notion of \emph{belief}, \emph{expectation} and \emph{goal}. In contrast, Lewis and Weigert~\cite{lewis1985trust} discuss the notion of experience within a collective social group. In this setting, \emph{an observation becomes an evidence which leads to expectation}. This is the fundamental approach to trust that we follow in this paper.

In the area of computing and information technology, the notion of trust is used in different areas including networking, security, artificial intelligence, human computer interaction and e-commerce, just to name a few~\cite{101007-1239199}~\cite{SICARI2015146}. There have been many works that discuss trust management models. It can be noted that the \emph{trust score} can be captured in various ways and based on the score computation, different design choices can be made~\cite{YAN2014120}~\cite{trust-is-relative-2012}. The core concept of trust in computing systems is derived from the social sciences. In general, trust can be decomposed into various aspects e.g., device trust, entity trust and data trust~\cite{7881789}. Cho et~al.~\cite{5604602} define trust as \emph{a subjective belief about whether an agent will exhibit behaviour reliably in a particular context with potential risks}. The \emph{belief} is based on learning from past experience to maximize utility (or minimize risk).

The notion of subjective belief has been well studied in other works as well~\cite{JOSANG2007618, mui2002computational}. The formal definition of trust in social sciences extends to computer systems. For instance, Kimery et~al.~\cite{kimery2002third} discuss trust for online systems. Trust is defined as \textit{``a consumer's willingness to accept vulnerability in an online transaction based on their positive expectations regarding an e-retailer's future behaviors''}.  This definition denotes the predicted behaviours of the users in an online system. Similarly, Cynthia et~al.~\cite{CORRITORE2003737} define trust as \textit{``an attitude of confident expectation in an online situation of risk that one's vulnerabilities will not be exploited''}. In both cases, a user computes a belief to measure the reliability of services to maximize utility or minimize risk. In this case, the user employs computed belief (i.e., trust) to resolve choices between multiple services.

Artz and Gil~\cite{ARTZ200758} define trust based on reputation. Reputation is defined as \textit{``an assessment based on the history of interactions or observations, either directly with the evaluator (personal experience) or as reported by others (recommendations or third party verification)''}. Since we assume evidence-based trust, reputation building is closely related to our work, but we take a more objective approach. That is, reputations are linked to proven interactions and are objectively verifiable by all involved parties.

In~\cite{Josang-2001-565981}, one of the earliest important works on trust in ICT, a trust-based framework is defined based on subjective logic. This framework explicitly considers uncertainty during trust evaluation. The arguments of subjective logic are called opinions. An opinion is denoted as ${^A}\omega_X$, where $A$ is the source of the said opinion and $X$ represents the state of the variable in which the particular opinion exists. Each opinion in subjective logic is equivalent to the binomial opinion of the beta distribution function. These functions are employed to calculate the success rate of an event based on the previous knowledge of interactions. 

More recently, considerable attention has been devoted to the management of trust frameworks.  For example, Sharma et~al.~\cite{7882970} present a generic framework to manage trust considering qualitative and quantitative parameters. In this framework, the trust management encompasses multiple phases that are dedicated to different activities. No implementation is given to support the framework. Moreover, no contextual information is taken into consideration for trust computation. Aied et~al.~\cite{BENSAIED2013351} propose a  context-aware and multi-service trust management system that uses past experiences of interactions when calculating the trust value. The proposed solution takes into consideration the various resource capabilities of the interacting entities in a heterogeneous environment in order to establish a community of trusted elements that respect the objectives of the operation of a set of collaborative services. The trust management system is controlled and governed by a trust manager. A simulation-based experimental study is conducted to show the performance of the proposed system in managing trust and enforcing collaboration between the nodes. This approach employs a centralised trust manager, with all the known disadvantages of that approach. The proposal is also limited in scope, being specifically targeted at co-operative services. Further, it is difficult to see how the approach could be extended to more general services. Wang et~al.~\cite{wang2013distributed} present a distributed trust management mechanism for large-scale systems. The authors extract three basic elements namely, service (i.e., defines the role of the trust management system), decision-making (i.e., making a decision to deliver a service) and self-organising (i.e., selecting a decision by the trust management system), of trust management and then, based on a service model, a trust management framework is established for the system. In this work, a fuzzy-based approach is employed for trust value calculation. In~\cite{6263792}, each device evaluates trust for a limited set of devices that holds the same interest using both direct observations and indirect recommendations. In~\cite{6513398}, the authors extend the work presented in~\cite{6263792} in particular focused on Community of Interest (CoI) based social systems where nodes can dynamically join and leave the system at any time. The major contributions of this proposal are the dynamic adaptation and scalability management  of the trust management protocol in the context of highly large and scalable systems. The dynamic adaptability property is demonstrated by showing that a new node in the community can quickly build its trust relationship with other nodes with desirable accuracy and convergence behaviour. To demonstrate the scalability, an efficient storage management strategy is introduced keeping the view of limited storage space of resource-constrained nodes. In a storage space, each node can keep the corresponding trust information towards a subset of nodes according to their interest and storage space. A similar concept of trust management system is presented by Chen et~al. in~\cite{7097037}. However, our work differs from these in that we provide evidence based trust with verifiable interactions and each interaction is linked to a feedback. 

The emergence of blockchain technology has been an important recent development in Information Technology~\cite{kiviat2015beyond}. Blockchain was originally developed as a distributed ledger whose absolute truth is voted through a consensus algorithm. Thus, it removes the reliance on a trusted third-party. In a blockchain, the ledger is held as a sequence of blocks recording transactions. The sequence is linked through cryptographic hashes where the last block contains the hash of its parent~\cite{8029379}. The consensus algorithm ensures that every node in the network agrees with the state of the ledger and by extension all valid transactions stored in the blocks. This information is reliable and transactions are traceable to the underlying accounts. Being distributed, auditable, robust and scalable makes the blockchain an important framework for other distributed applications~\cite{7945805}. 

There has been considerable recent interest in supporting trust through the use of blockchain technology. For instance, Pietro et~al.~\cite{di2018blockchain} present a blockchain-based trust system for large-scale dynamic systems like the Internet of Things (IoT). In this work, a distributed trust management model is introduced that takes advantage of existing trust domains and helps bridge them to provide an end-to-end trust. Trust between the entities does not rely upon a common root of trust. Blockchain technology is used to link the established trust between interacting entities into longer trust chains. However, it does not employ blockchain in the management of interaction evidence. Zhang and Zhou~\cite{9096382} propose the use of blockchains to hold trust items of nodes, which includes reputation and trust values and trust-relevant data e.g., running, waiting and processing time. However, this is part of their proposed future work, so no implementation is given and no detail is provided about the manner in which this would be achieved.

Other contributions have also proposed storing trust-related information in a blockchain. This includes information about the interacting entities or the messages they exchange. For example, Huang et~al.~\cite{9016397} store information about the requestors and service providers for offloaded computations in a parked vehicle assisted fog computing environment. The results of computations are uploaded to the chain where miners check that the computations meet the task requirements. This contribution also uses a financial-based approach in assigning rewards for performing tasks. However, the evidence in the blockchain is not directly used in the computation of trust values. Lu et~al.~\cite{8455893} discuss a blockchain-based anonymous reputation system that is used for trust management in VANETs (vehicular ad-hoc networks). However, in that proposal blockchains are used to record the public key certificates of the validated entities and the revocation of these certificates. Trust related information is held by a central entity (in this case, a law enforcement agency), not held on the blockchain. Kang et~al.~\cite{8489897} discuss the use of blockchain to support data sharing in vehicular edge computing. Reputation is employed to decide with which other vehicles it is safe to share data and subjective logic is used in the calculation of the reputation values. Smart contracts manage the storage and data and meta-data on the blockchains. Each vehicle calculates trust values separately based on their own interactions and the blockchains, and the information stored in them, play no role in this calculation. Similarly, in Wu and Ansari~\cite{9222063}, the blockchain is not used directly for trust management. Instead, the blockchain is used to secure the access control. Each device stores and maintains its own trust-related information.

Some contributions have made use of blockchains for the storage and management of trust-related information.  For example, the final, calculated, trust values can be stored on the blockchain. For example, Shala et~al.~\cite{SHALA2019100058} for machine to machine application services. Test agent evaluate the trust level (1-5) of services and, after computing the trust level, store this information on a public blockchain network. The aim of Kim et~al.~\cite{8936389} is to enhance trust between nodes in wireless sensor networks and eliminate malicious nodes from the network. Trust in this proposal is based on a number of factors, e.g., closeness, honesty, intimacy and frequency of interaction and is computed by each node for each other node. Once trust values are computed they are placed in a blockchain by a Base Station. Again, no other trust information in placed on the blockchain. Another similar approach, from a trust point of view, is Yang et~al.~\cite{8358773} where, again, final calculated trust values are stored in the blockchain, in this case trust values are calculated by road-side units on vehicles in vehicular networks. In none of these examples is other trust-related information, e.g., actual evidence, stored on the blockchain

There are other contributions which have made more extensive use of the possibilities offered by blockchain technology to support trust management. One of the earliest examples of this is Dennis and Owen~\cite{7412073}. Their contribution addressed peer-to-peer networks, where for a each transaction a record whether the transaction was positive or negative was added to the blockchain. There was, however, unlike our system, no guarantee that such an addition represented an actual transaction. Also, trust values themselves were not added to the blockchain. Another example is Zhao et~al.~\cite{zhao2020trustblock}, which places both trust values and data related to node behaviour in the blockchain.  The data related to node behaviour is quite detailed. While this may be suitable for the application are of that contribution (Software Defined Networks), the presence of detailed information about node performance in a public blockchain is not advisable for general situations, as it risks security and privacy breaches. In contrast, our proposal places evidence of the interaction occurring on the blockchain, not sensitive data about the content of those interactions. A similar example is Boussard et~al.~\cite{8845126}, which also addresses SDN networks. Again, sensitive data is stored on the blockchain, by observers of the behaviour of nodes. This is not directly relevant to systems which involve actual interaction and relies on implicit trust in the observers. They do, however, have the concept of an analyzer, which uses the data on the blockchain to derive trust values and is related to our concept of trust providers. 

Dedeoglu et~al. \cite{1033607743360822} propose a trust system for sensor nodes involving blockchain. The actual data from observations is not stored on the blockchain, instead transactions recording the collection of data and the reputation of nodes is stored on the blockchain. However, as with the previous contribution, this relies on implicit in the data trust modules. A further example is Adnan et~al.~\cite{khan2019secure}, where trust is used to evaluate messages sent between members of a VANET. Both the messages and the calculated trust values are placed on blockchains. However, the messages in this system concern events observed by the nodes and the trust calculation is based on the number of other nodes observing the same event.  It is unclear how the approach could be generalised to systems where events are not shared in this manner. There is also no checks made that a node has actually observed an event. It could simply be repeating a report from another node to improve its trustworthiness. The system also requires four separate blockchains in its implementation, which is a significant overhead. A similar approach is taken by Zhao et~al.~\cite{zhao2020trustblock} which addresses trust in nodes in a SDN. While both evidence and trust values are stored in the blockchain, the evidence relies on readily observable behaviour of the nodes, such as switch throughput.  This approach is not readily adaptable to situations involving evidence that derives from private interactions between two parties in the network. Similarly for Yan et~al.~\cite{yan2021social}, where the nodes are trusted to generate the evidence and there is no check that an interaction actually occurs.

As can be seen from the foregoing discussion, there has been considerable interest in employing blockchain techniques to enhance trust management. However, our proposal is unique in both giving an assurance that an interaction has actually occurred, without relying on either readily observable behaviour and/or a privileged central component, and in allowing multiple trust providers to access the evidence and provide separately calculated trust values. This leverages the well-known financial applications of blockchain technology and allows the implementation of trust-as-a-service.

%%%
\section{Conclusion}
\label{conclusion}
In this paper, we have developed a formal framework for trust systems where evidences are backed up by provable interaction records. The approach is simple yet powerful enough to capture various instances of scoring ranging from simple averaging to more complicated functions that account for the type of services and temporal signatures on the interactions and feedbacks. %{\color{red} SP: Do we need this line -- We have shown that the formal framework is sound as long as two fundamental properties are guaranteed.} 

We designed a blockchain platform that implements our framework (using Ethereum blockchain). In particular, we have shown that the properties provided by blockchains are crucial to guarantee the two fundamental properties required for the framework to be sound. %The platform has been implemented using the Ethereum blockchain. 
The performance analysis of the prototype implementation supports the feasibility of our approach. 
We also provided a detailed discussion on the general benefits that our approach contributes to improving the resilience of trust systems to attacks, in particular to hostile trust providers. While our system inherits the same mitigation capabilities as centralised trust systems, it provides further protections which come from verifiable interaction and the inherent financial cost for all blockchain transactions. 

The present state of our work has some limitations. For instance, we take the universal set of evidence as the ground truth in the sense that it is void of uncertainty. However, the atomic review ratings themselves will have some inherent uncertainty as it is next to impossible that all users are very sure of the ratings they give for the interactions they had. We also did not investigate the notion of trust update which is crucial for implementing trust systems efficiently. We leave these for future work.

\section*{References}
\bibliography{mybibfile}

\begin{thebibliography}{10}
\expandafter\ifx\csname url\endcsname\relax
  \def\url#1{\texttt{#1}}\fi
\expandafter\ifx\csname urlprefix\endcsname\relax\def\urlprefix{URL }\fi
\expandafter\ifx\csname href\endcsname\relax
  \def\href#1#2{#2} \def\path#1{#1}\fi

\bibitem{101007-1239199}
C.~Jensen, \href{https://link.springer.com/chapter/10.1007/978-3-662-43813-8_1}{The importance of trust in computer security}, in: J.~Zhou,
  N.~Gal-Oz, J.~Zhang, E.~Gudes (Eds.), Trust Management VIII, Springer Berlin
  Heidelberg, Berlin, Heidelberg, 2014, pp. 1--12.
\url{https://doi.org/10.1007/978-3-662-43813-8_1}

\bibitem{ALTAF2019}
A.~Altaf, H.~Abbas, F.~Iqbal, A.~Derhab,
  \href{http://www.sciencedirect.com/science/article/pii/S1084804519300839}{Trust
  models of internet of smart things: A survey, open issues and future
  directions}, Journal of Network and Computer Applications, 2019, Vol.~137, pp. 93--111. 
\url{http://www.sciencedirect.com/science/article/pii/S1084804519300839}

\bibitem{truong2016survey}
N.~Truong, U.~Jayasinghe, T.~Um, G.~Lee, \href{https://www.researchgate.net/publication/316042146_A_Survey_on_Trust_Computation_in_the_Internet_of_Things}{A survey on trust computation}
  in the internet of things, The Journal of Korean Institute of Communications
  and Information Sciences (JKICS) 33~(2) (2016) 10--27.
\url{https://www.koreascience.or.kr/article/JAKO201608949924661.page}

\bibitem{JAFARIAN2020107254}
B.~Jafarian, N.~Yazdani, M.~{Sayad Haghighi},
  \href{https://www.sciencedirect.com/science/article/pii/S1389128619316743}{Discrimination-aware
  trust management for social internet of things}, Computer Networks 178, 2020.
\url{https://www.sciencedirect.com/science/article/pii/S1389128619316743}

\bibitem{GHASEMPOURI2019571}
S.~Ghasempouri, B.~{Tork Ladani},
  \href{http://www.sciencedirect.com/science/article/pii/S0167739X18326013}{Modeling
  trust and reputation systems in hostile environments}, Future Generation
  Computer Systems 99 (2019) 571--592.
\url{http://www.sciencedirect.com/science/article/pii/S0167739X18326013}

\bibitem{CHEN2021107952}
G.~Chen, F.~Zeng, J.~Zhang, T.~Lu, J.~Shen, W.~Shu,
  \href{https://www.sciencedirect.com/science/article/pii/S138912862100089X}{An
  adaptive trust model based on recommendation filtering algorithm for the
  internet of things systems}, Computer Networks 190 (2021) 107952.
\url{https://www.sciencedirect.com/science/article/pii/S138912862100089X}

\bibitem{1024434627}
T.~Noor, Q.~Sheng, \href{https://link.springer.com/chapter/10.1007/978-3-642-24434-6_27}{Trust as a service: A framework for trust management
  in cloud environments}, in: A.~Bouguettaya, M.~Hauswirth, L.~Liu (Eds.), Web
  Information System Engineering -- WISE 2011, Springer Berlin Heidelberg,
  Berlin, Heidelberg, 2011, pp. 314--321.
\url{https://doi.org/10.1007/978-3-642-24434-6_27}

\bibitem{HASSAN2019512}
M.~Hassan, M.~Rehmani, J.~Chen,
  \href{http://www.sciencedirect.com/science/article/pii/S0167739X18326542}{Privacy
  preservation in blockchain based iot systems: Integration issues, prospects,
  challenges, and future research directions}, Future Generation Computer
  Systems 97 (2019) 512--529.
\url{http://www.sciencedirect.com/science/article/pii/S0167739X18326542}

\bibitem{8386853}
Y.~{Zhang}, S.~{Kasahara}, Y.~{Shen}, X.~{Jiang}, J.~{Wan},  \href{https://ieeexplore.ieee.org/document/8386853}{Smart
  contract-based access control for the internet of things}, IEEE Internet of
  Things Journal 6~(2) (2019) 1594--1605.
\url{https://doi.org/10.1109/JIOT.2018.2847705}

\bibitem{BROTSIS2021108005}
S.~Brotsis, K.~Limniotis, G.~Bendiab, N.~Kolokotronis, S.~Shiaeles,
  \href{https://www.sciencedirect.com/science/article/pii/S1389128621001225}{On
  the suitability of blockchain platforms for iot applications: Architectures,
  security, privacy, and performance}, Computer Networks 191 (2021) 108005.
\url{https://www.sciencedirect.com/science/article/pii/S1389128621001225}

\bibitem{8894097}
S.~{Pal}, T.~{Rabehaja}, A.~{Hill}, M.~{Hitchens}, V.~{Varadharajan}, \href{https://ieeexplore.ieee.org/abstract/document/8894097}{On the
  integration of blockchain to the internet of things for enabling access right
  delegation}, IEEE Internet of Things Journal 7~(4) (2020) 2630--2639.
\url{https://doi.org/10.1109/JIOT.2019.2952141}

\bibitem{DBLP:journals-5313}
S.~Prajapati, S.~Changder, A.~Sarkar,
  \href{http://arxiv.org/abs/1304.5313}{Trust management model for cloud
  computing environment}, CoRR abs/1304.5313.
\newblock \href {http://arxiv.org/abs/1304.5313} {\path{arXiv:1304.5313}}.
\newline\urlprefix\url{http://arxiv.org/abs/1304.5313}

\bibitem{BELLOUSMAN2018143}
A.~Usman, J.~Gutierrez,
  \href{http://www.sciencedirect.com/science/article/pii/S1570870518304906}{Toward
  trust based protocols in a pervasive and mobile computing environment: A
  survey}, Ad Hoc Networks 81 (2018) 143 -- 159.
\url{http://www.sciencedirect.com/science/article/pii/S1570870518304906}

\bibitem{BEYNON200037}
M.~Beynon, B.~Curry, P.~Morgan,
  \href{http://www.sciencedirect.com/science/article/pii/S030504839900033X}{The
  dempster–shafer theory of evidence: an alternative approach to
  multicriteria decision modelling}, Omega 28~(1) (2000) 37 -- 50.
\url{http://www.sciencedirect.com/science/article/pii/S030504839900033X}

\bibitem{dempster2008upper}
A.~Dempster, \href{https://link.springer.com/chapter/10.1007/978-3-540-44792-4_3}{Upper and lower probabilities induced by a multivalued mapping},
  Vol.~38, Ann. Math. Statist., 1967, pp. 325--339.
\url{https://doi.org/10.1007/978-3-540-44792-4_3}

\bibitem{Josang-2001-565981}
A.~J{\o}sang, \href{http://dx.doi.org/10.1142/S0218488501000831}{A logic for
  uncertain probabilities}, Int. J. Uncertain. Fuzziness Knowl.-Based Syst.
  9~(3) (2001) 279--311.
\url{http://dx.doi.org/10.1142/S0218488501000831}

\bibitem{6838647}
I.~{Chen}, J.~{Guo}, \href{https://ieeexplore.ieee.org/document/6838647}{Dynamic hierarchical trust management of mobile groups and
  its application to misbehaving node detection}, in: 2014 IEEE 28th
  International Conference on Advanced Information Networking and Applications,
  2014, pp. 49--56.
\url{http://dx.doi.org/10.1109/AINA.2014.13}

\bibitem{6513398}
F.~Bao, I.~Chen, J.~Guo, \href{http://dx.doi.org/10.1109/ISADS.2013.6513398}{Scalable, adaptive and survivable trust management for
  community of interest based internet of things systems}, in: 2013 IEEE
  Eleventh International Symposium on Autonomous Decentralized Systems (ISADS),
  2013, pp. 1--7.
\url{http://dx.doi.org/10.1109/ISADS.2013.6513398}

\bibitem{JonkerTreur}
C.~Jonker, J.~Treur, \href{https://link.springer.com/chapter/10.1007/3-540-48437-X_18}{Formal analysis of models for the dynamics of trust
  based on experiences}, in: European workshop on modelling autonomous agents in
  a multi-agent world, Springer, 1999, pp. 221--231.
\url{https://doi.org/10.1007/3-540-48437-X_18}

\bibitem{dempster-upper}
A.~Dempster, \href{https://projecteuclid.org/journals/annals-of-mathematical-statistics/volume-38/issue-2/Upper-and-Lower-Probabilities-Induced-by-a-Multivalued-Mapping/10.1214/aoms/1177698950.full}{Upper and lower probabilities induced by a multivalued mapping},
  Ann. Math. Statist 38 (1967) 325--339.
\url{http://dx.doi.org/10.1214/aoms/1177698950}

\bibitem{8752021}
S.~{Pal}, T.~{Rabehaja}, M.~{Hitchens}, V.~{Varadharajan}, A.~{Hill}, \href{http://dx.doi.org/10.1109/TII.2019.2925898}{On the
  design of a flexible delegation model for the internet of things using
  blockchain}, IEEE Transactions on Industrial Informatics 16~(5) (2020)
  3521--3530.
\url{doi:10.1109/TII.2019.2925898}.

\bibitem{yan2021social}
Z.~Yan, L.~Peng, W.~Feng, L.~T. Yang, \href{https://dl.acm.org/doi/abs/10.1145/3419102}{Social-chain: Decentralized trust
  evaluation based on blockchain in pervasive social networking}, ACM
  Transactions on Internet Technology (TOIT) 21~(1) (2021) 1--28.
\url{https://doi.org/10.1145/3419102}

\bibitem{GUO20171}
J.~Guo, I.~Chen, J.~Tsai,
  \href{http://www.sciencedirect.com/science/article/pii/S0140366416304959}{A
  survey of trust computation models for service management in internet of
  things systems}, Computer Communications 97 (2017) 1--14.
\url{https://doi.org/10.1016/j.comcom.2016.10.012}

\bibitem{4068036}
Y.~{Sun}, Z.~{Han}, W.~{Yu}, K.~{Ray Liu},\href{https://ieeexplore.ieee.org/document/4068036}{ Attacks on trust evaluation in
  distributed networks}, in: The 40th Annual Conference on Information Sciences
  and Systems, 2006, pp. 1461--1466.
\url{https://doi.org/10.1109/CISS.2006.286695}

\bibitem{cloudarmor}
T.~Noor, Q.~Sheng, A.~Ngu, A.~Alfazi, J.~Law, \href{http://dx.doi.org/10.1145/2505515.2508204}{Cloud armor: a platform for
  credibility-based trust management of cloud services}, 2013, pp. 2509--2512.
\url{https://doi:10.1145/2505515.2508204}

\bibitem{LI2020841}
X.~Li, P.~Jiang, T.~Chen, X.~Luo, Q.~Wen,
  \href{http://www.sciencedirect.com/science/article/pii/S0167739X17318332}{A
  survey on the security of blockchain systems}, Future Generation Computer
  Systems 107 (2020) 841--853. 
\url{http://www.sciencedirect.com/science/article/pii/S0167739X17318332}

\bibitem{SI20191028}
H.~Si, C.~Sun, Y.~Li, H.~Qiao, L.~Shi,
  \href{http://www.sciencedirect.com/science/article/pii/S0167739X19312725}{Iot
  information sharing security mechanism based on blockchain technology},
  Future Generation Computer Systems 101 (2019) 1028--1040.
\url{http://www.sciencedirect.com/science/article/pii/S0167739X19312725}

\bibitem{jayasinghe2018trust}
U.~Jayasinghe, \href{http://web.science.mq.edu.au/~yanwang/Maryam-MRes-Thesis-final.pdf}{Trust evaluation in the iot environment}, Ph.D. thesis, Liverpool
  John Moores University (2018).
\url{http://web.science.mq.edu.au/~yanwang/Maryam-MRes-Thesis-final.pdf}

\bibitem{QURESHI2021107647}
K.~Qureshi, G.~Jeon, F.~Piccialli,
  \href{https://www.sciencedirect.com/science/article/pii/S1389128620312664}{Anomaly
  detection and trust authority in artificial intelligence and cloud
  computing}, Computer Networks 184 (2021) 107647.
\url{https://www.sciencedirect.com/science/article/pii/S1389128620312664}

\bibitem{WU2021108004}
Y.~Wu, Z.~Wang, Y.~Ma, V.~Leung,
  \href{https://www.sciencedirect.com/science/article/pii/S1389128621001213}{Deep
  reinforcement learning for blockchain in industrial iot: A survey}, Computer
  Networks 191 (2021) 108004.
\url{https://www.sciencedirect.com/science/article/pii/S1389128621001213}

\bibitem{arai2009defining}
K.~Arai, \href{https://www.jstor.org/stable/43296226?seq=1}{Defining trust using expected utility theory, Hitotsubashi Journal of
  Economics} (2009) 205--224.
\url{https://www.jstor.org/stable/43296226?seq=1}

\bibitem{Cho-2815595}
J.~Cho, K.~Chan, S.~Adali, \href{http://doi.acm.org/10.1145/2815595}{A
  survey on trust modeling}, ACM Comput. Surv. 48~(2) (2015) 28:1--28:40.
\url{http://doi.acm.org/10.1145/2815595}

\bibitem{MOUSA201549}
H.~Mousa, S.~Mokhtar, O.~Hasan, O.~Younes, M.~Hadhoud, L.~Brunie,
  \href{https://www.sciencedirect.com/science/article/pii/S1389128615002340}{Trust
  management and reputation systems in mobile participatory sensing
  applications: A survey}, Computer Networks 90 (2015) 49--73.
\url{https://www.sciencedirect.com/science/article/pii/S1389128615002340}

\bibitem{evans2009psychology}
A.~Evans, J.~Krueger, \href{https://onlinelibrary.wiley.com/doi/abs/10.1111/j.1751-9004.2009.00232.x}{The psychology (and economics) of trust, Social and
  Personality Psychology Compass}, 3~(6) (2009) 1003--1017.
\url{https://doi.org/10.1111/j.1751-9004.2009.00232.x}

\bibitem{rousseau1998not}
D.~Rousseau, S.~Sitkin, R.~Burt, C.~Camerer, \href{https://journals.aom.org/doi/10.5465/amr.1998.926617}{Not so different after
  all: A cross-discipline view of trust}, Academy of management review,23~(3)
  (1998) 393--404.
\url{https://doi.org/10.5465/amr.1998.926617}

\bibitem{deutsch1962cooperation}
M.~Deutsch, \href{https://psycnet.apa.org/record/1964-01869-002}{Cooperation and trust: Some theoretical notes},
in M.~Jones (Ed.), Nebraska Symposium on Motivation, 1962, pp. 275--320. Univer. Nebraska Press.
\url{https://psycnet.apa.org/record/1964-01869-002}

\bibitem{101007-3}
D.~Harrison~McKnight, N.~Chervany, \href{https://link.springer.com/chapter/10.1007/3-540-45547-7_3}{Trust and distrust definitions: One bite
  at a time}, in: R.~Falcone, M.~Singh, Y.~Tan (Eds.), Trust in
  Cyber-societies, Springer Berlin Heidelberg, Berlin, Heidelberg, 2001, pp.
  27--54.
\url{https://doi.org/10.1007/3-540-45547-7_3}

\bibitem{101007978-3}
A.~J{\o}sang, S.~Presti, \href{https://link.springer.com/chapter/10.1007/978-3-540-24747-0_11}{Analysing the relationship between risk and trust},
  in: C.~Jensen, S.~Poslad, T.~Dimitrakos (Eds.), Trust Management, Springer
  Berlin Heidelberg, Berlin, Heidelberg, 2004, pp. 135--145.
\url{https://doi.org/10.1007/978-3-540-24747-0_11}

\bibitem{lewis1985trust}
J.~Lewis, A.~Weigert, \href{https://www.jstor.org/stable/2578601}{Trust as a social reality}, Social forces 63~(4) (1985)
  967--985.
\url{https://www.jstor.org/stable/2578601}

\bibitem{SICARI2015146}
S.~Sicari, A.~Rizzardi, L.~Grieco, A.~Coen-Porisini,
  \href{https://www.sciencedirect.com/science/article/pii/S1389128614003971}{Security,
  privacy and trust in internet of things: The road ahead}, Computer Networks
  76 (2015) 146--164.
\url{https://www.sciencedirect.com/science/article/pii/S1389128614003971}

\bibitem{YAN2014120}
Z.~Yan, P.~Zhang, A.~Vasilakos,
  \href{http://www.sciencedirect.com/science/article/pii/S1084804514000575}{A
  survey on trust management for internet of things}, Journal of Network and
  Computer Applications 42 (2014) 120--134.
\url{http://www.sciencedirect.com/science/article/pii/S1084804514000575}

\bibitem{trust-is-relative-2012}
L.~Fritsch, A.-Groven, T.~Schulz, \href{https://link.springer.com/chapter/10.1007/978-3-642-31479-7_46}{On the internet of things, trust is
  relative}, in: R.~Wichert, K.~Van~Laerhoven, J.~Gelissen (Eds.), Constructing
  Ambient Intelligence, Springer Berlin Heidelberg, Berlin, Heidelberg, 2012,
  pp. 267--273.
\url{https://doi.org/10.1007/978-3-642-31479-7_46}

\bibitem{7881789}
A.~{Arabsorkhi}, M.~{Sayad Haghighi}, R.~{Ghorbanloo}, \href{http://dx.doi.org/10.1109/ISTEL.2016.7881789}{A conceptual trust model
  for the internet of things interactions}, in: The 8th International Symposium
  on Telecommunications (IST), 2016, pp. 89--93.
\url{https://doi:10.1109/ISTEL.2016.7881789}

\bibitem{5604602}
J.~{Cho}, A.~{Swami}, I.~{Chen}, \href{http://dx.doi.org/10.1109/SURV.2011.092110.00088}{A survey on trust management for mobile ad hoc
  networks}, IEEE Communications Surveys Tutorials 13~(4) (2011) 562--583.
\url{https://doi:10.1109/SURV.2011.092110.00088}

\bibitem{JOSANG2007618}
A.~Jøsang, R.~Ismail, C.~Boyd,
  \href{http://www.sciencedirect.com/science/article/pii/S0167923605000849}{A
  survey of trust and reputation systems for online service provision},
  Decision Support Systems 43~(2) (2007) 618--644, emerging Issues in
  Collaborative Commerce.
\url{https://doi.org/10.1016/j.dss.2005.05.019}

\bibitem{mui2002computational}
L.~Mui, \href{https://dspace.mit.edu/handle/1721.1/87343}{Computational models of trust and reputation: Agents, evolutionary
  games, and social networks}, Ph.D. thesis, Massachusetts Institute of
  Technology (2002).
\url{https://dspace.mit.edu/handle/1721.1/87343}

\bibitem{kimery2002third}
K.~Kimery, M.~McCord, \href{https://aisel.aisnet.org/cgi/viewcontent.cgi?referer=https://www.google.com/&httpsredir=1&article=1179&context=jitta}{Third party assurances: mapping the road to trust in
  eretailing}, Journal of Information Technology Theory and Application (JITTA)
  4~(2) (2002).
\url{https://aisel.aisnet.org/cgi/viewcontent.cgi?referer=https://www.google.com/&httpsredir=1&article=1179&context=jitta}

\bibitem{CORRITORE2003737}
C.~Corritore, B.~Kracher, S.~Wiedenbeck,
  \href{http://www.sciencedirect.com/science/article/pii/S1071581903000417}{On-line
  trust: concepts, evolving themes, a model}, International Journal of
  Human-Computer Studies 58~(6) (2003) 737--758, trust and Technology.
\url{https://doi.org/10.1016/S1071-5819(03)00041-7}

\bibitem{ARTZ200758}
D.~Artz, Y.~Gil,
  \href{http://www.sciencedirect.com/science/article/pii/S1570826807000133}{A
  survey of trust in computer science and the semantic web}, Journal of Web
  Semantics 5~(2) (2007) 58--71, software Engineering and the Semantic Web.
\url{https://doi.org/10.1016/j.websem.2007.03.002}

\bibitem{7882970}
A.~{Sharma}, E.~{Pilli}, A.~{Mazumdar}, M.~{Govil}, \href{http://dx.doi.org/10.1109/ETCT.2016.7882970}{A framework to
  manage trust in internet of things}, in: The International Conference on
  Emerging Trends in Communication Technologies (ETCT), 2016, pp. 1--5.
\url{https://doi:10.1109/ETCT.2016.7882970}

\bibitem{BENSAIED2013351}
Y.~Saied, A.~Olivereau, D.~Zeghlache, M.~Laurent,
  \href{http://www.sciencedirect.com/science/article/pii/S0167404813001302}{Trust
  management system design for the internet of things: A context-aware and
  multi-service approach}, Computers \& Security 39 (2013) 351--365.
\url{https://doi.org/10.1016/j.cose.2013.09.001}

\bibitem{wang2013distributed}
J.~Wang, S.~Bin, Y.~Yu, X.~Niu, \href{https://www.scientific.net/AMM.347-350.2463}{Distributed trust management mechanism
  for the internet of things}, in: Applied Mechanics and Materials, Vol. 347,
  Trans Tech Publ, 2013, pp. 2463--2467.
\url{https://doi.org/10.4028/www.scientific.net/AMM.347-350.2463}

\bibitem{6263792}
F.~Bao, I.~Chen, \href{http://dx.doi.org/10.1109/WoWMoM.2012.6263792}{Trust management for the internet of things and its
  application to service composition}, in: IEEE International Symposium on
  a World of Wireless, Mobile and Multimedia Networks (WoWMoM), 2012, pp. 1--6.
\url{https://doi:10.1109/WoWMoM.2012.6263792}

\bibitem{7097037}
I.~Chen, F.~Bao, J.~Guo, \href{http://dx.doi.org/10.1109/TDSC.2015.2420552}{Trust-based service management for social internet of
  things systems}, IEEE Transactions on Dependable and Secure Computing 13~(6)
  (2016) 684--696.
\url{https://doi:10.1109/TDSC.2015.2420552}

\bibitem{kiviat2015beyond}
T.~Kiviat, \href{https://scholarship.law.duke.edu/dlj/vol65/iss3/4/}{Beyond bitcoin: Issues in regulating blockchain tranactions}, Duke
  LJ 65 (2015) 569.
\url{https://scholarship.law.duke.edu/dlj/vol65/iss3/4/}

\bibitem{8029379}
Z.~Zheng, S.~Xie, H.~Dai, X.~Chen, H.~Wang, \href{http://dx.doi.org/10.1109/BigDataCongress.2017.85}{An overview of blockchain
  technology: Architecture, consensus, and future trends}, in: IEEE
  International Congress on Big Data (BigData Congress), 2017, pp. 557--564.
\url{https://doi:10.1109/BigDataCongress.2017.85}

\bibitem{7945805}
M.~Conoscenti, A.~Vetrò, J.~Martin, \href{http://dx.doi.org/10.1109/AICCSA.2016.7945805}{Blockchain for the internet of
  things: A systematic literature review}, in: IEEE/ACS 13th International
  Conference of Computer Systems and Applications (AICCSA), 2016, pp. 1--6.
\url{https://doi:10.1109/AICCSA.2016.7945805}

\bibitem{di2018blockchain}
R.~Pietro, X.~Salleras, M.~Signorini, E.~Waisbard, \href{https://dl.acm.org/doi/abs/10.1145/3205977.3205993}{A blockchain-based trust
  system for the internet of things}, in: Proceedings of the 23nd ACM on
  Symposium on Access Control Models and Technologies, 2018, pp. 77--83.
\url{https://doi.org/10.1145/3205977.3205993}

\bibitem{9096382}
P.~{Zhang}, M.~{Zhou}, \href{http://dx.doi.org/10.1109/TCSS.2020.2990103}{Security and trust in blockchains: Architecture, key
  technologies, and open issues}, IEEE Transactions on Computational Social
  Systems 7~(3) (2020) 790--801.
\url{https://doi:10.1109/TCSS.2020.2990103}

\bibitem{9016397}
X.~{Huang}, D.~{Ye}, R.~{Yu}, L.~{Shu}, \href{http://dx.doi.org/10.1109/JAS.2020.1003039}{Securing parked vehicle assisted fog
  computing with blockchain and optimal smart contract design}, IEEE/CAA Journal
  of Automatica Sinica 7~(2) (2020) 426--441.
\url{https://doi:10.1109/JAS.2020.1003039}

\bibitem{8455893}
Z.~{Lu}, Q.~{Wang}, G.~{Qu}, Z.~{Liu}, \href{https://ieeexplore.ieee.org/document/8455893}{Bars: A blockchain-based anonymous
  reputation system for trust management in vanets}, in: The 17th IEEE
  International Conference On Trust, Security And Privacy In Computing And
  Communications/ 12th IEEE International Conference On Big Data Science And
  Engineering (TrustCom/BigDataSE), 2018, pp. 98--103.
\url{https://doi:10.1109/TrustCom/BigDataSE.2018.00025}

\bibitem{8489897}
J.~{Kang}, R.~{Yu}, X.~{Huang}, M.~{Wu}, S.~{Maharjan}, S.~{Xie}, Y.~{Zhang},
  \href{http://dx.doi.org/10.1109/JIOT.2018.2875542}{Blockchain for secure and efficient data sharing in vehicular edge computing
  and networks}, IEEE Internet of Things Journal 6~(3) (2019) 4660--4670.
\\url{https://doi:10.1109/JIOT.2018.2875542}

\bibitem{9222063}
D.~{Wu}, N.~{Ansari}, \href{http://dx.doi.org/10.1109/JIOT.2020.3030689}{A trust-evaluation-enhanced blockchain-secured industrial
  iot system}, IEEE Internet of Things Journal 8~(7) (2021) 5510--5517.
\url{https://doi:10.1109/JIOT.2020.3030689}

\bibitem{SHALA2019100058}
B.~Shala, U.~Trick, A.~Lehmann, B.~Ghita, S.~Shiaeles,
  \href{https://www.sciencedirect.com/science/article/pii/S2542660519301234}{Novel
  trust consensus protocol and blockchain-based trust evaluation system for m2m
  application services}, Internet of Things 7 (2019) 100058.
\url{https://doi.org/10.1016/j.iot.2019.100058}

\bibitem{8936389}
T.~Kim, R.~Goyat, M.~Rai, G.~Kumar, W.~Buchanan, R.~Saha, R.~Thomas, \href{http://dx.doi.org/10.1109/ACCESS.2019.2960609}{A
  novel trust evaluation process for secure localization using a decentralized
  blockchain in wireless sensor networks}, IEEE Access 7 (2019) 184133--184144.
\url{https://doi:10.1109/ACCESS.2019.2960609}

\bibitem{8358773}
Z.~Yang, K.~Yang, L.~Lei, K.~Zheng, V.~Leung, \href {http://dx.doi.org/10.1109/JIOT.2018.2836144}{Blockchain-based
  decentralized trust management in vehicular networks}, IEEE Internet of Things
  Journal 6~(2) (2019) 1495--1505.
\url{https://doi:10.1109/JIOT.2018.2836144}

\bibitem{7412073}
R.~Dennis, G.~Owen, \href {http://dx.doi.org/10.1109/ICITST.2015.7412073}{Rep on the block: A next generation reputation system based
  on the blockchain}, in: 10th International Conference for Internet
  Technology and Secured Transactions (ICITST), 2015, pp. 131--138.
\url{https://doi:10.1109/ICITST.2015.7412073}

\bibitem{zhao2020trustblock}
B.~Zhao, Y.~Liu, X.~Li, J.~Li, J.~Zou, \href{https://journals.plos.org/plosone/article?id=10.1371/journal.pone.0228844}{Trustblock: An adaptive trust evaluation
  of sdn network nodes based on double-layer blockchain}, PloS one 15~(3) (2020)
  e0228844.
\url{https://doi.org/10.1371/journal.pone.0228844}

\bibitem{8845126}
M.~Boussard, S.~Papillon, P.~Peloso, M.~Signorini, E.~Waisbard, \href{http://dx.doi.org/10.1109/INFCOMW.2019.8845126}{Steward:sdn and
  blockchain-based trust evaluation for automated risk management on iot
  devices}, in: IEEE INFOCOM - IEEE Conference on Computer Communications
  Workshops (INFOCOM WKSHPS), 2019, pp. 841--846.
\url{https://doi:10.1109/INFCOMW.2019.8845126}

\bibitem{1033607743360822}
V.~Dedeoglu, R.~Jurdak, G.~D. Putra, A.~Dorri, S.~S. Kanhere,
  \href{https://doi.org/10.1145/3360774.3360822}{A trust architecture for
  blockchain in iot}, in: Proceedings of the 16th EAI International Conference
  on Mobile and Ubiquitous Systems: Computing, Networking and Services,
  MobiQuitous'19, Association for Computing Machinery, New York, NY, USA,
  2019, p. 190–199.
\url{https://doi.org/10.1145/3360774.3360822}

\bibitem{khan2019secure}
A.~Khan, K.~Balan, Y.~Javed, S.~Tarmizi, J.~Abdullah, \href{https://www.mdpi.com/1424-8220/19/22/4954}{Secure trust-based
  blockchain architecture to prevent attacks in vanet}, Sensors 19~(22) (2019)
  4954.
\url{https://doi.org/10.3390/s19224954}

\end{thebibliography}

\end{document}